\documentclass[sigconf]{acmart}
\usepackage{graphicx}
\usepackage{wrapfig}
\usepackage{caption}
\usepackage{blindtext}
\usepackage{bbm} 
\usepackage[capitalise]{cleveref}
\usepackage{enumitem}
\usepackage{array}
\usepackage{xcolor}
\usepackage{soul}
\usepackage{subcaption}
\usepackage{tikz}
\usepackage{seqsplit}
\usepackage{placeins}
\usepackage{circledsteps}
\usepackage{balance}
\usepackage{tikz}

\newcommand{\btri}{$\tikz\fill(0,-.2em) -- (.25em,.333em) -- (.5em,-.2em) -- cycle;$}

\usepackage{scalerel}

\definecolor{alex}{HTML}{045A8D}
\definecolor{alexLPIPS}{HTML}{2B8CBE}
\definecolor{alexSTY}{HTML}{74A9CF}
\definecolor{squeeze}{HTML}{feb24c} 
\definecolor{squeezeLPIPS}{HTML}{ffd97f} 
\definecolor{vgg}{HTML}{006d2c}
\definecolor{vggLPIPS}{HTML}{2ca25f}
\definecolor{vggSTY}{HTML}{66c2a4}
\definecolor{resnet18}{HTML}{810F7C}
\definecolor{resnet50}{HTML}{8856a7}
\definecolor{resnet50STY}{HTML}{8c96c6}
\definecolor{efftnetB0}{HTML}{b30000}
\definecolor{efftnetB5}{HTML}{e34a33}
\definecolor{l5}{HTML}{E41A1C}
\definecolor{l9}{HTML}{377EB8}
\definecolor{sa}{HTML}{4DAF4A}
\definecolor{td}{HTML}{984EA3}
\definecolor{tm}{HTML}{FF7F00}

\DeclareRobustCommand{\ColCircle}[1]{
\begin{tikzpicture}
\filldraw[fill={#1},draw={#1}] circle (2pt);
\end{tikzpicture}
}

\newcommand{\task}[2]{{\textcolor{#1}{\textsf{#2}}}} 

\AtBeginDocument{%
  }

\copyrightyear{2025}
\acmYear{2025}
\setcopyright{cc}
\setcctype{by}
\acmConference[CHI '25]{CHI Conference on Human Factors in Computing Systems}{April 26-May 1, 2025}{Yokohama, Japan}
\acmBooktitle{CHI Conference on Human Factors in Computing Systems (CHI '25), April 26-May 1, 2025, Yokohama, Japan}\acmDOI{10.1145/3706598.3713955}
\acmISBN{979-8-4007-1394-1/25/04}

\begin{document}

\title[Seeing Eye to AI?]{Seeing Eye to AI? Applying Deep-Feature-Based Similarity Metrics to Information Visualization}

\author{Sheng Long}
\email{shenglong@u.northwestern.edu}
\affiliation{%
  \institution{Northwestern University}
  \institution{ }
  \city{Evanston}
  \state{Illinois}
  \country{USA}
}
\orcid{0009-0000-9752-5898}

\author{Angelos Chatzimparmpas}
\email{a.chatzimparmpas@uu.nl}
\affiliation{
  \institution{Utrecht University}
  \institution{ }
  \city{Utrecht}
  \country{Netherlands}
}
\orcid{0000-0002-9079-2376}

\author{Emma Alexander}
\email{ealexander@northwestern.edu}
\affiliation{%
  \institution{Northwestern University}
  \institution{ }
  \city{Evanston}
  \state{Illinois}
  \country{USA}
}
\orcid{0000-0003-0159-3111}

\author{Matthew Kay}
\email{mjskay@northwestern.edu}
\affiliation{%
  \institution{Northwestern University}
  \institution{ }
  \city{Evanston}
  \state{Illinois}
  \country{USA}
}
\orcid{0000-0001-9446-0419}

\author{Jessica Hullman}
\email{jhullman@northwestern.edu}
\affiliation{%
  \institution{Northwestern University}
  \institution{ }
  \city{Evanston}
  \state{Illinois}
  \country{USA}
}
\orcid{0000-0001-6826-3550}


\begin{abstract}
  Judging the similarity of visualizations is crucial to various applications, such as visualization-based search and visualization recommendation systems. Recent studies show deep-feature-based similarity metrics correlate well with perceptual judgments of image similarity and serve as effective loss functions for tasks like image super-resolution and style transfer. We explore the application of such metrics to judgments of visualization similarity. We extend a similarity metric using five ML architectures and three pre-trained weight sets. We replicate results from previous crowdsourced studies on scatterplot and visual channel similarity perception. Notably, our metric using pre-trained ImageNet weights outperformed gradient-descent tuned MS-SSIM, a multi-scale similarity metric based on luminance, contrast, and structure. Our work contributes to understanding how deep-feature-based metrics can enhance similarity assessments in visualization, potentially improving visual analysis tools and techniques. Supplementary materials are available at \url{https://osf.io/dj2ms/}.
\end{abstract}



\maketitle
\section{Introduction} 

Similarity is a fundamental construct in human cognition, underlying numerous mental processes, including categorization, reasoning, and decision-making \cite{medinRespectsSimilarity1993}. Human-perceived similarity has been studied extensively in psychology and cognitive science \cite{kriegeskorte2008representational, roads2024modeling} and has been leveraged in applications such as image retrieval \cite{guo2002learning} and human-in-the-loop categorization \cite{wah2015learning, roads2017improving} in recent years \cite{roads2021enriching}. 
Recent studies in computer vision have shown that deep-feature-based similarity metrics correlate well with perceptual judgments of image similarity ~\cite{zhangUnreasonableEffectivenessDeep2018} and serve as effective loss functions for tasks like image super-resolution ~\cite{johnson2016perceptual} and style transfer ~\cite{gatysNeuralAlgorithmArtistic2015}.
The success of these applications in computer vision raises an intriguing question: \textit{Can similar approaches be effectively applied to the domain of information visualization?} 

Judging visualization similarity is crucial for various applications, ranging from visual search \cite{wolfe2020visual} and visualization recommendation \cite{zeng2021evaluation} to automated design and sequencing~\cite{demiralp2014learning, chen2023does}.
Knowing whether a visualization is \textit{robustly discriminable} across a range of datasets, i.e., whether it consistently maintains clear visual distinctions across a range of datasets, is crucial for effective communication and design. The idea that perceived visual structure should correspond with structure in the underlying data is recurring in multiple works, such as in the principle of visual-data correspondence by Kindlmann and Scheidegger \cite{kindlmann2014algebraic}, in the principle of congruence by Tversky et al. \cite{tversky1977features}, in similar principles proposed in multi-view visualization~\cite{qu2016evaluating,qu2017keeping}, and in the visual embedding model by Demiralp et al. \cite{demiralpVisualEmbeddingModel2014}. 
However, such knowledge is challenging to obtain at design time and costly to evaluate empirically \cite{veras2019discriminability}. 

This paper investigates the use of deep-feature-based similarity metrics to approximate similarity perception in information visualization. 
We offer two primary research contributions.

\begin{enumerate}
    \item While deep neural networks are widely used in visualization for various tasks ~\cite{wu2021ai4vis, wang2021survey}, our work is the first to implement a domain-independent transfer learning technique from computer vision \cite{zhangUnreasonableEffectivenessDeep2018, kumarBetterImageNetClassifiers2022} to an information visualization setting. We extended prior work on deep-feature-based similarity metrics using weights trained on \textit{Stylized ImageNet} \cite{geirhos2018imagenet}, a modified ImageNet-1K dataset where images are artistically stylized while preserving their original content and labels. Given Stylized ImageNets' increased object detection performance, we hypothesized it would benefit information visualization tasks.
    Our results suggest that transfer learning from computer vision can reduce the time and resources needed to develop effective models when applied with caution to information visualization, potentially opening up new avenues for cross-domain research and applications. However, we did not observe any significant performance differences between models trained on Stylized ImageNet versus ImageNet-1K.
    \item We assess the benefits and limitations of applying deep-feature-based similarity metrics in information visualization by conceptually replicating and comparing our results to two crowd-sourced studies that collected similarity judgments \cite{demiralp2014learning, veras2019discriminability}. These studies are well-suited for exploring visualization similarity as they offer diverse datasets and methodologies, allowing us to systematically examine how deep-feature-based similarity measures perform across different visualization contexts and analytical approaches. 
    
    \begin{enumerate}
        \item Our first replication (\cref{sec:veras}) reveals that when using certain deep learning (DL) networks, \textit{which have not been trained on scatterplots}, deep-feature-based similarity metrics achieve better clustering alignment with human judgments of scatterplot similarity than traditional computer vision metrics whose parameters are optimized on the set of scatterplots.
        \item Our second replication (\cref{sec:demiralp}) reveals that for visual channels like color and shape, deep-feature-based similarity metrics struggle to capture what humans perceive to be similar. However, these metrics perform well when assessing the visual channel of size.
    \end{enumerate}
\end{enumerate}

We speculate that the strong performance of deep-feature-based similarity metrics in capturing scatterplot similarity (\cref{sec:veras}) is due to their ability to capture patterns in spatial distributions of visual elements. This suggests potential applications to other visualization types that also encode data through spatial arrangements. These metrics struggle to capture judgments of color and shape similarity (\cref{sec:demiralp}), which may be due to humans making these judgments based on factors beyond spatial features, such as cultural association. The metrics' better performance with size judgments likely stems from its more direct relationship to spatial distribution. These results highlight both opportunities and limitations in applying deep-feature-based similarity metrics trained on natural scenes to information visualization. 
\section{Background}
\label{sec:background}

\begin{figure*}[!htbp]
    \centering 
    \includegraphics[width=0.75\textwidth]{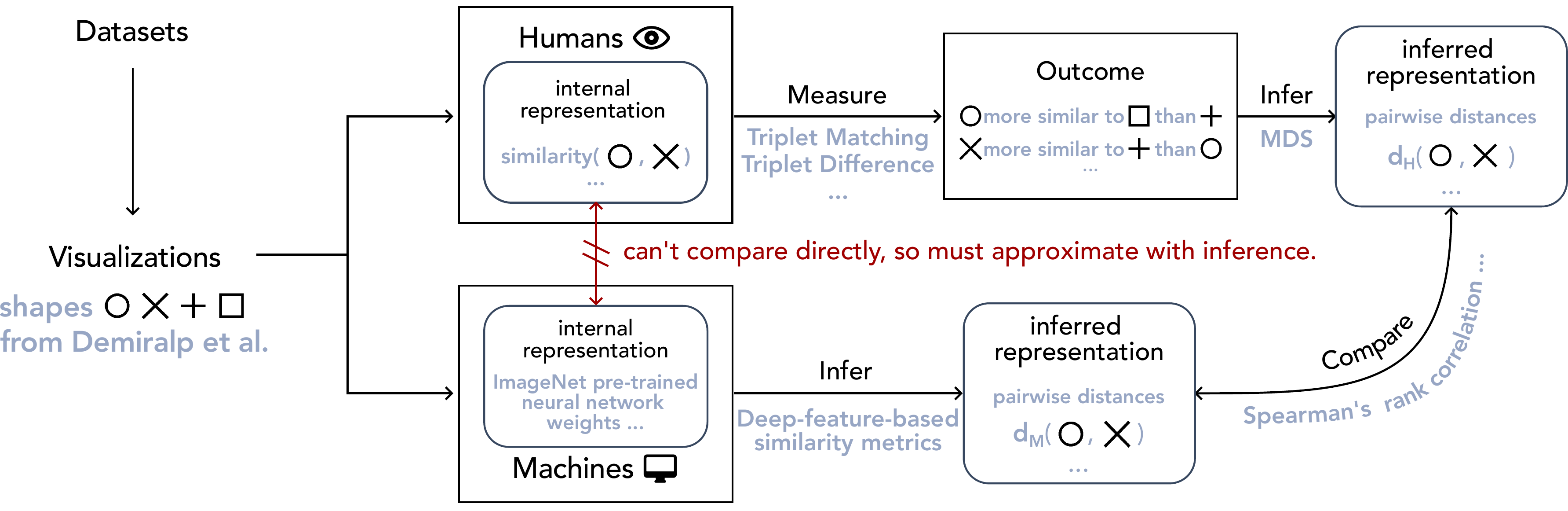}
    \caption{A general framework for comparing two information-processing systems (e.g., humans vs. machines), inspired by the framework proposed by Sucholutsky et al. \cite{sucholutsky2023getting}. Framework uses Demiralp et al.'s experiment on the perception of shape similarity as one concrete example \cite{demiralp2014learning}. }
    \Description{A diagram describing the general framework this paper uses to compare humans' internal representation with internal representations of deep neural networks. Uses the shape experiment from Demiralp et al. as an example.}
 \label{fig:teaser}   
\end{figure*}

\subsection{Perceptual Similarity Metrics}

Computer vision has long studied the problem of measuring the quality of an image for downstream tasks such as lossless compression. Due to the costly nature of running experiments to collect human data, significant research effort has been devoted to deriving an \textit{objective} metric that can \textit{approximate human perception}. For example, if two images appear \textit{similar} to the human eye, then their ``distance'', measured by this metric, should be close to 0. 

Traditional metrics, such as Manhattan $\mathcal{L}_1$, Euclidean $\mathcal{L}_2$, Mean Squared Error (MSE), and Peak-Signal-to-Noise-Ratio (PSNR), rely on \textit{point-wise} pixel differences, which offers computational ease but poorly matches perceived visual quality \cite{wangImageQualityAssessment2004, eskicioglu1995image, girod1993s}. Subsequently developed patch-based metrics rely on the hypothesis that the human visual system is highly adapted for extracting \textit{structural} information. One such metric is Similarity Index Measure (SSIM) \cite{wangImageQualityAssessment2004}, which compares ``local patterns of pixel intensities that have been normalized for luminance and contrast''. One popular extension of SSIM is \textit{Multi-Scale SSIM} (MS-SSIM) \cite{wang2003multiscale}, which extends SSIM by evaluating image similarity at \textit{multiple scales}, capturing both local and global structural information. Nevertheless, SSIM and MS-SSIM often do not capture the nuances of human vision when more structural ambiguity is present \cite{sampat2009complex} and fall short for more complex downstream tasks, such as image generation and image synthesis. 

More recently, researchers have leveraged deep neural networks (DNN) for tasks beyond classification to develop metrics such as LPIPS~\cite{zhangUnreasonableEffectivenessDeep2018}, PieAPP~\cite{prashnaniPieAPPPerceptualImageError2018} and DISTS~\cite{dingImageQualityAssessment2020} that utilize \textit{deep features}, i.e., features obtained through deep-learning architectures trained on the ImageNet dataset \cite{deng2009imagenet}. One of the primary motivations for utilizing these deep-feature-based metrics is that they can transform pixel representations to a space that is more \textit{perceptually uniform}~\cite{dingImageQualityAssessment2020}. They have also demonstrated excellent empirical performance for downstream tasks such as image super-resolution~\cite{johnson2016perceptual} and style transfer~\cite{gatysNeuralAlgorithmArtistic2015}. However, their application to the perceptual similarity of information visualization remains unexplored, perhaps due to the uncertainty about whether features learned from natural scenes can effectively transfer to the often abstract and stylistic domain of information visualization --- a gap this work addresses. 

\subsubsection{Collecting Human Similarity Judgments} 
\label{sec:collect_human_judgment}

Regardless of the underlying mechanics, the usefulness of similarity metrics is ultimately determined by how well they align with human judgment. Traditional Image Quality Assessment (IQA) databases, such as LIVE~\cite{sheikh2003image}, CSIQ~\cite{larson2010most}, and TID2013~\cite{ponomarenko2015image}, collect \textit{Mean Opinion Scores} by asking participants to provide subjective ratings for pairwise images on a specified numerical scale (e.g., Likert scales).  More recent IQA databases, such as BAPPS~\cite{zhangUnreasonableEffectivenessDeep2018}, use two-alternative forced choice (2AFC) experiments to collect similarity judgment on a dataset of images applied with low-level distortions. Beyond low-level perceptual similarity, the THINGS database collected behavioral odd-one-out similarity judgments~\cite{hebart2019things, hebart2020revealing} on everyday objects to generate interpretable object dimensions predictive of behavior and similarity, and the NIGHTS dataset~\cite{fu2023dreamsim} collected 2AFC judgment on diffusion-synthesized images perturbed along various dimensions. 

Aside from the computer vision/machine learning databases mentioned above, information visualization studies have also collected data on similarity judgments (albeit on a much smaller scale). Demiralp et al.~\cite{demiralp2014learning} collected similarity judgment via five different tasks on different visual channels, such as color, size, and shape. Pandey et al.~\cite{pandey2016towards} asked participants to group scatterplot thumbnails and compared their results to scatterplot diagnostics (scagnostics). Ma et al.~\cite{ma2018scatternet} collected similarity judgment on scatterplots and trained a deep neural network, \texttt{ScatterNet}, to learn features that capture the human perception of scatterplot similarity. Beyond scatterplots, Gogolou et al.~\cite{gogolou2018comparing} collected similarity perception of time-series plots by asking participants to select among four options which time-series chart appears most similar to the reference chart. 
Others have collected judgments of visualization similarity in the form of transition costs when moving from one visualization to the next in a sequence~\cite{hullman2013deeper,kim2017graphscape}. In this paper, we conduct conceptual replications of the experiments by Demiralp et al.~\cite{demiralp2014learning} and Veras and Collins~\cite{veras2019discriminability} (that uses Pandey et al.'s data~\cite{pandey2016towards}). 

\subsection{Comparing Human and Machine Behavior} 

Given the remarkable performance of DNNs at image classification, research across different fields --- computer vision, cognitive science, neural science, to name a few ---  has sought to answer \textbf{whether DNNs are good models of the human visual system}. To answer this question requires \textit{comparing} DNNs to the human visual system. Existing research has done this both at the \textit{mechanistic} level (i.e., comparing internal representations) and \textit{functional/behavioral} level (i.e., comparing outcomes). See Sucholutsky et al.~\cite{sucholutsky2023getting} for an overview of comparing and aligning representations across different information processing systems.

\subsubsection{Functional Similarity}

At the \textbf{outcome} level (\cref{fig:teaser}), much work has focused on comparing and modeling classification performance~\cite{roads2021enriching}. Researchers have found that small perturbations that are imperceptible to humans can dramatically affect model classification decisions~\cite{goodfellow2014explaining}, and that texture and local image features drive classifiers~\cite{baker2018deep, geirhos2018imagenet}, whereas humans are more strongly influenced by Gestalt shape~\cite{attarian2020transforming}. Beyond comparing classification accuracies, other work~\cite{geirhosAccuracyQuantifyingTrialbytrial2020} has looked at additional outcome measures such as trial-by-trial error consistency.

In information visualization, Haehn et al. ~\cite{haehnEvaluatingGraphicalPerception2019} trained four different convolutional neural network (CNN) architectures on five different visualization tasks to reproduce Cleveland and McGill's graphical perception experiments~\cite{cleveland1984graphical}. They found that when compared to human performance baselines, CNNs are not ``currently a good model for human graphical perception'' --- while CNNs performed better than humans in elementary perceptual tasks (e.g., estimating quantities from visual marks), humans significantly out-performed CNNs in visual relation tasks (e.g., comparing bar lengths). Yang et al. ~\cite{yang2023can} trained DNNs on \textit{human correlation judgments} in scatterplots across three studies and found that a subset of their trained DL architectures (e.g., VGG-19) has comparable accuracy to the best-performing regression analyses in prior research. The main difference between our work and Haehn et al.~\cite{haehnEvaluatingGraphicalPerception2019} is that we are not training network parameters on any domain-specific dataset but instead rely on pre-trained ImageNet and Stylized ImageNet weights. The main difference between our work and Yang et al.~\cite{yang2023can} is that we utilize DL architectures that are trained on natural images for classification and not on human judgment. Our metrics are therefore \textit{information-visualization-domain-independent} and can be applied directly to any pair of visual stimuli,\footnote{... so long as their spatial dimensions exceed $64 \times 64$. Details in~\Cref{sec:exp-overview}.} whereas the trained weights and architectures of Haehn et al. ~\cite{haehnEvaluatingGraphicalPerception2019} and Yang et al. ~\cite{haehnEvaluatingGraphicalPerception2019} may face challenges in generalizing to contexts beyond their specific training domains.

\subsubsection{Representational Similarity}

At the \textbf{internal representation} level (\cref{fig:teaser}), numerous techniques exist, and the most popular one is \textit{Representational Similarity Analysis} (RSA), a multivariate technique introduced by Kriegeskorte et al.~\cite{kriegeskorte2008representational}. RSA compares different representations of the same set of stimuli, such as neural activity patterns, computational model outputs, and behavioral data, by computing similarity matrices for each representation and then comparing these matrices to assess how well different representations align with each other. Using RSA, Khaligh-Razavi and Kriegeskorte~\cite{khaligh2014deep} compared the representational geometry of object recognition in the human brain with that of various computer vision models, and they found that a deep convolutional network performed best in both categorization and explaining inferior temporal cortex representations. For more on calculating similarity in psychological space, see Roads and Love~\cite{roads2024modeling} for an overview. 

\subsubsection{Texture vs. Shape}

Most DNN models that perform well on Brain-Score~\cite{schrimpf2018brain} and other prediction metrics \textit{do not} rely on global shape when classifying objects, contrary to the fundamental conclusion from psychological research that humans largely rely on shape when identifying objects~\cite{bowers2023deep}. DNNs (such as the CORnet-S model~\cite{kubilius2018cornet}, one of the best models of human vision) largely classify objects based on \textit{texture} and \textit{local shape features}~\cite{wichmann2023deep}. 

Prior work~\cite{geirhos2018imagenet} showed that ML architectures can learn shape-based representations when trained on \textit{Stylized ImageNet}, a version of ImageNet where images have their visual appearances transformed by applying artistic styles. ML architectures trained on the Stylized ImageNet have better object detection performance and robustness towards a wide range of image distortions. To the best of our knowledge, existing deep-feature-based similarity metrics have not examined the performance of deep features extracted from networks trained on stylized images. We investigate whether the robust performance of ML architectures trained on Stylized ImageNet can transfer to the domain of information visualization, where humans tend to make judgments based on shape by extending the existing implementation of deep-feature-based similarity metrics, with details in~\cref{sec:neural_network_weights}. 

\section{Research Goals and Experiment Overview}
\label{sec:exp-overview}

Our ultimate goal is to assess the benefits and limitations of deep-feature-based perceptual similarity metrics for information visualization. For this goal, we conceptually replicated two experiments: 

\begin{enumerate}
    \item Veras and Collins \cite{veras2019discriminability} (using data from Pandey et al. \cite{pandey2016towards}) 
    \item Demiralp et al. \cite{demiralp2014learning}
\end{enumerate}

Both studies provide high-quality, crowd-sourced human similarity judgments gathered on stimuli with varying visualization complexity. Together, these studies cover both holistic similarity judgments of data visualizations and fine-grained perceptual comparisons of visual encodings. 

Our replication experiments follow a consistent procedure across both studies: assign pairwise distances between stimuli from the original studies using deep-feature-based similarity metrics, analyze the results using the same procedures as the original papers, and compare our results to the analyzed human judgment results in the original studies (\cref{fig:teaser}). Due to the inability of deep-feature-based similarity metrics (specifically \texttt{AlexNet}) to process images with dimensions smaller than $63 \times 63$, we generated new stimuli of size $224 \times 224$ when replicating Demiralp et al.~\cite{demiralp2014learning} in \cref{sec:demiralp}.

\begin{figure*}[!ht]
    \centering
    \includegraphics[width=\linewidth]{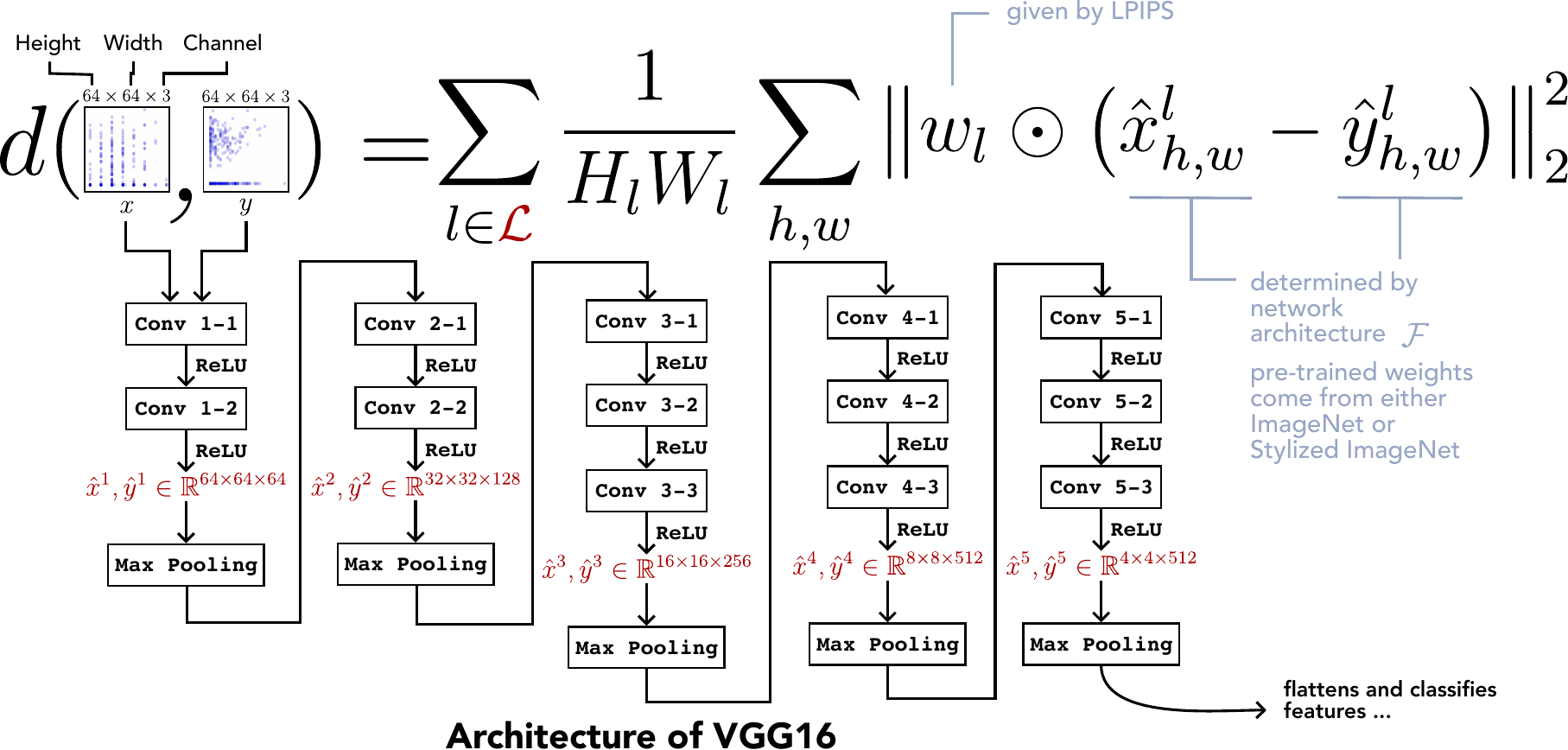}
    \caption{Diagram showing how the deep-feature-based perceptual similarity metric calculates ``perceptual distance'', using \texttt{VGG16} as the DL network $\mathcal{F}$.}
    \label{fig:deep-learning-pipeline}
    \Description{A diagram that walks through the formal definition for extracting deep features when we pass two images into a deep-featured-based similarity function.}
\end{figure*}

\section{Implementing Deep-Feature-Based Perceptual Similarity Metrics}
\label{sec:implement}

We implement deep-feature-based perceptual similarity metrics that use ML architectures and three pre-trained weights. Our implementation primarily follows the implementations of Zhang et al. \cite{zhangUnreasonableEffectivenessDeep2018} and Kumar et al. \cite{kumarBetterImageNetClassifiers2022}. We focus our attention on supervised models trained on the ImageNet-1k dataset \cite{deng2009imagenet}, which contains 1, 281,167 training images, 50,000 validation images, and 100,000 test images.

Given two images $x, y \in \mathbb{R}^{H \times W \times C}$, where $H, W, C$ correspond to \textbf{h}eight, \textbf{w}idth, and \textbf{c}hannel, and a network $\mathcal{F}$, the \textit{``perceptual'' distance} between $x$ and $y$ is the weighted sum of squared differences between feature activations of $x$ and $y$ across multiple layers and spatial positions. Formally, it is defined as 
\begin{align}
    d(x, y) = \sum_{l \in \mathcal{L}} \frac{1}{H_l W_l} \sum_{h, w} \left \Vert w_l \odot \left(\hat{x}_{h, w}^l - \hat{y}_{h, w}^l \right) \right \Vert_2^2 
    \label{eq:distance_function}
\end{align}
where $\mathcal{L}$ is the set of layers in network  $\mathcal{F}$ from which features are extracted\footnote{the set of extraction layers $\mathcal{L}$ is architecture-specific; details of  $\mathcal{L}$ in \Cref{sec:extraction_points}}, $\hat{x}^l, \hat{y}^l \in \mathbb{R}^{H_l \times \, W_l \times \, C_l}$ are the unit-normalized\footnote{at the channel dimension} deep feature maps extracted by $\mathcal{F}$ at layer $l$, and vector $w_l \in \mathbb{R}^{C_l}$ is a channel-wise \textit{scaling vector} for the difference between unit-normalized feature maps $\hat{x}^l$ and $\hat{y}^l$ at spatial location $(h, w)$. 

Using out-of-the-box pre-trained weights from ImageNet or Stylized ImageNet corresponds to $w_l = \mathbf{1}_{C_l} \; \forall l \in \mathcal{L}$, where $\mathbf{1}_{C_l}$ is a vector of ones with length $C_l$, the number of channels at layer $l$. In other words, we use the extracted deep features \textit{without any scaling or linear calibration}. 
Using LPIPS (version 0.1) \footnote{i.e., ``\texttt{lin}'' in Zhang et al. \cite{zhangUnreasonableEffectivenessDeep2018}} corresponds to using ImageNet pre-trained weights \textit{scaled by} parameters $w_l$ learned from BAPPS \cite{zhangUnreasonableEffectivenessDeep2018}, a dataset of 16K patches derived by applying exclusively low-level distortions to the MIT-Adobe 5k dataset \cite{bychkovsky2011learning} for training and the RAISE1k data \cite{dang2015raise} for validation. In other words, $w^l, l \in \mathcal{L}$ constitute a ``perceptual calibration'' of a few parameters in an existing feature space \cite{zhangUnreasonableEffectivenessDeep2018}. Zhang et al. \cite{zhangUnreasonableEffectivenessDeep2018} showed that LPIPS achieve better performance than traditional metrics (e.g., $L_2$, SSIM~\cite{wang2003multiscale}) and correlates well with human similarity judgments.

Our selection of DNN architectures is motivated by both theoretical considerations and empirical evidence from computer vision and prior work on training and predicting human similarity judgments~\cite{yang2023can}. We build upon Zhang et al.'s ~\cite{zhangUnreasonableEffectivenessDeep2018} finding that deep features from CNNs trained on natural images correlate strongly with human perceptual judgments. Our architectural choices span from SqueezeNet~\cite{iandola2016squeezenet} and AlexNet ~\cite{krizhevsky2012imagenet} to more complex Residual Nets ~\cite{he2016deep} and Efficient Nets~\cite{tan2019efficientnet}, selected to systematically investigate how different levels of hierarchical feature extraction affect visualization similarity judgments. This range of architectures, covering ImageNet-1K classification accuracies from 56.522\% (AlexNet) to 83.444\% (EfficientNet B5), allows us to test whether improved natural image classification correlates with better visualization similarity assessment. 

While newer architectures like Vision Transformers ~\cite{dosovitskiyImageWorth16x162021} exist, we prioritized CNNs for their architectural flexibility and relevance to our context: (1) CNNs allow processing of different input sizes, unlike Vision Transformers which require fixed input dimensions; (2) CNNs provide scale-invariant feature detection through their hierarchical structure; and (3) CNNs enable direct comparability with validated perceptual similarity metrics like LPIPS~\cite{zhangUnreasonableEffectivenessDeep2018}. This flexibility is particularly important as our implementation builds on LPIPS, which has been rigorously validated against human perceptual judgments using the Berkeley-Adobe Perceptual Patch Similarity (BAPPS) dataset at $64\times64$ resolution. Importantly, our approach is architecture-agnostic and can be readily extended to any CNN architecture that provides hierarchical feature maps, enabling future comparisons to additional architectures.

\begin{table}[!htbp]
    \centering
    \caption{Network architectures and weights}
    \begin{tabular}{l l}
        \toprule
        \textbf{Network Architecture} & \textbf{Weight} \\
        \midrule 
        AlexNet \cite{krizhevsky2012imagenet} & ImageNet, Stylized ImageNet, LPIPS \\ 
        VGG16 \cite{simonyan2014very} & ImageNet, Stylized ImageNet, LPIPS\\ 
        SqueezeNet \cite{iandola2016squeezenet} & ImageNet, LPIPS\\ 
        ResNet18 \cite{he2016deep} & ImageNet \\ 
        ResNet50 \cite{he2016deep} & ImageNet, Stylized ImageNet\\ 
        EfficientNetB0 \cite{tan2019efficientnet} & ImageNet  \\ 
        EfficientNetB5 \cite{tan2019efficientnet} & ImageNet \\ 
        \bottomrule
    \end{tabular}
    \label{tab:architecture_and_weights}
\end{table}

We use \texttt{pytorch} \cite{paszke2019pytorch}, \texttt{torchvision }weights\footnote{\url{https://pytorch.org/vision/stable/models.html}}, and Zhang et al.'s implementation of \texttt{LPIPS} \cite{zhangUnreasonableEffectivenessDeep2018} at \url{https://github.com/richzhang/PerceptualSimilarity}. We use the following \texttt{R} packages for computing and presenting our results: \texttt{reticulate}~\cite{ushey}, \texttt{tidyverse}~\cite{wickham2019welcome}, \texttt{ggplot2}~\cite{wickham2011ggplot2}, \texttt{aricode}~\cite{chiquet2020package}, \texttt{MatrixCorrelation}~\cite{matrixcorrelation}, \texttt{ggtext}~\cite{wilke2020ggtext}, and \texttt{ggpubr}~\cite{ggpubr}. 

\subsection{Pre-processing}

We follow the same image pre-processing steps of Zhang et al. \cite{zhangUnreasonableEffectivenessDeep2018}.\footnote{Namely, we convert the input stimuli into sRGB color space, resize them to $64 \times 64 \times 3$, and normalize them by the same mean and standard deviation coefficients used by Zhang et al.~\cite{zhangUnreasonableEffectivenessDeep2018}, which are the same mean and standard deviation for ImageNet-1K except adjusted for inputs ranging from [-1, 1] instead of [0, 1].} This approach differs from the standard $224 \times 224 \times 3$ resolution that ImageNet-1K is typically trained on. For completeness, we have also run the same experiments using the standard resolution with details provided in \Cref{sec:process_224}. While the overall trend remains similar, using the standard resolution leads to slightly worse performance than with a lower resolution of $64 \times 64 \times 3$. This finding suggests that the lower resolution may be more suitable for our specific task. By choosing a smaller size, we focus on the low-level aspects of perceptual similarity to mitigate the effect of ``differing respects of similarity''~\cite{medinRespectsSimilarity1993} that may be influenced by high-level semantics~\cite{zhangUnreasonableEffectivenessDeep2018}. Furthermore, the lower resolution yields significant improvements in computational efficiency, most notably reducing the computing time for \texttt{VGG16} from approximately three hours to merely 30 minutes.\footnote{Experiments were conducted on a Dell XPS 15 (2019) with Intel Core i7-8750H @ 2.20GHz, 16GB RAM, NVIDIA GeForce GTX 1050 Ti with Max-Q Design, running Microsoft Windows 11 Pro.} 

Prior research has shown that pre-processing images by transforming them into different color spaces yielded different image classification results~\cite{gowda2019colornet}. We verify that all images in ImageNet-1K and BAPPS are in sRGB space, and we transform all input images into sRGB format before processing.  
\section[Replicating Veras and Collins]{Replicating Veras and Collins~\cite{veras2019discriminability}}
\label{sec:veras}

\subsection{Overview of Original Paper}

\begin{figure}[!htbp]
    \centering
    \includegraphics[width=\linewidth]{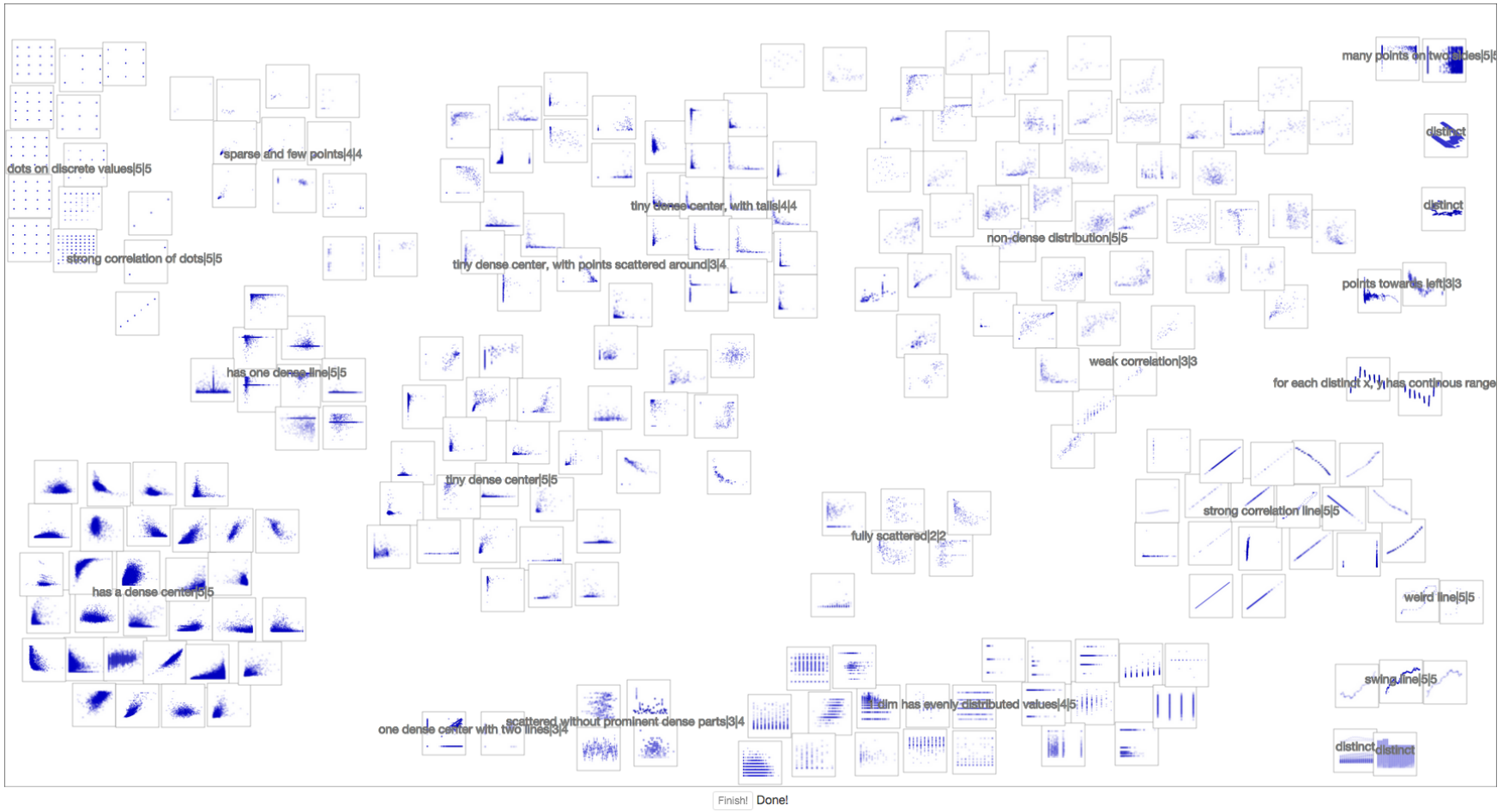}
    \caption{Screenshot obtained from the supplementary materials of Pandey et al.~\cite{pandey2016towards} showing Participant 1's final screen after completing the naming phase, with group descriptions and their corresponding easiness and confidence scores (separated by pipe characters).}
    \Description{Screenshot obtained from the supplementary materials of Pandey et al. showing Participant 1's final screen after completing the naming phase, with group descriptions and their corresponding easiness and confidence scores (separated by pipe characters).}
    \label{fig:pandey_example}
\end{figure}

Veras and Collins~\cite{veras2019discriminability} investigated the application of a well-established computer vision metric, the Multi-Scale Structural Similarity Index (MS-SSIM)~\cite{wang2003multiscale}, to assess the \textit{discriminability} of a data visualization across a variety of datasets. In this context, discriminability refers to ``the average perceptual distance between the corresponding visualizations'' for a collection of datasets~\cite{veras2019discriminability}. Their research proposed using perceptual similarity metrics to evaluate and rank competing visual encodings based on their discriminability, thereby informing visualization selection for specific data distributions.

Veras and Collins first used gradient descent to optimize MS-SSIM parameters using stimuli from a study of scatterplot similarity judgments by Pandey et al.~\cite{pandey2016towards}. With this tuned MS-SSIM, they calculated pairwise distances between scatterplots and derived grouping labels using hierarchical clustering. To validate their approach, they compared their group labels against the consensus group labels from human similarity judgments in Pandey et al.'s original study~\cite{pandey2016towards}.  

\subsubsection{Data, Visual Stimuli, Task}

Veras and Collins~\cite{veras2019discriminability} based their research on data from Pandey et al.'s study~\cite{pandey2016towards}, which investigated the subjective perception of similarity in scatterplots. Pandey et al.~\cite{pandey2016towards} conducted a study with 18 participants from scientific backgrounds, asking them to group similar scatterplot thumbnails and provide confidence and easiness ratings for their groupings (see~\cref{fig:pandey_example}). The participants were asked to group a total of 247 scatterplots. These human-generated groupings served as a benchmark for Veras and Collins~\cite{veras2019discriminability} to validate the performance of their tuned MS-SSIM. Data from Pandey et al. is provided at \url{https://github.com/nyuvis/scatter-plot-similarity}. 

\subsubsection{Similarity Measure}

Given that different participants might group the same pair of scatterplots into two different groups, Pandey et al.~\cite{pandey2016towards} calculated the \textit{consensus distance} between plots $i$ and $j$ as follows: 
\begin{align}
    d_{i,j} = \frac{1}{N} \sum_{k=1}^N \left ( 1- \frac{c_{i,j}}{\min\{c_i, c_j\} } \right )_k
\end{align}
where $N$ is the number of participants, $c_{i, j}$ is the number of clusters that contain both plots $i$ and $j$, and $c_i$ and $c_j$ are the number of clusters that contain the plots $i$ and $j$ respectively. The authors allowed participants to assign the same plot to multiple groups. These consensus distances formed a \textit{consensus perceptual distance matrix}, which was later clustered using hierarchical clustering to form groupings.  

Veras and Collins~\cite{veras2019discriminability} conceptually replicated Pandey et al.'s experiment~\cite{pandey2016towards} by using an alternative similarity measure to approximate the perceptual distance between scatterplots. They used the Multiscale-Structural Similarity Index (MS-SSIM)~\cite{wang2003multiscale}, an extension of the Structural Similarity Index (SSIM)~\cite{wangImageQualityAssessment2004} where the contrast and structural similarities are computed at $K$ image scales. For image pair $X, Y$, MS-SSIM is defined as 
\begin{align}
    \text{MS-SSIM}(X, Y) = l(x,y)^\alpha \prod_{i=1}^K c(x_i, y_i)^{\beta_i} s(x_i, y_i)^{\gamma_i}
\end{align}
where $l(\cdot)$ is the luminance similarity function, $c(\cdot)$ is the contrast similarity function, and $s(\cdot)$ is the structural similarity function. Veras and Collins set $\alpha = 1, \beta_i = \gamma_i = w_i$ for $i \in [K]$, where $K = 5$ is the number of scales. The weights $w_i$ can be interpreted as the relative importance of each image at scale $i$ in determining similarity. Veras and Collins~\cite{veras2019discriminability} employed an iterative process to adjust the scale weights $w_1, w_2, ..., w_5$ in order to minimize the discrepancy between the similarity scores calculated by MS-SSIM and a set of empirical human judgments on scatterplots. 

\subsubsection{Performance Evaluation}

Veras and Collins used pairwise distances calculated by fine-tuned MS-SSIM to construct clustering labels for the scatterplots and compared these clustering labels against empirical judgments using four established \textit{cluster quality measures}\footnote{Cluster quality measures are traditionally used to qualify the agreement between two independent label assignments on the same dataset.} --- adjusted mutual information (AMI), normalized mutual information (NMI), Rand index (RI), and adjusted Rand index (ARI). Based on information theory, NMI and AMI quantify the shared information between two clusterings, with NMI scaling mutual information to $[0,1]$, while AMI correcting for chance. They are robust to differences in cluster numbers and sizes. RI measures the percentage of correct decisions made by the clustering algorithm, considering all possible pairs of points, while ARI adjusts RI for chance, providing a more reliable comparison between clusterings. All measures except RI range from 0 (random clustering) to 1 (perfect clustering relative to the ground truth).

\subsection{Implementation}

We extend Veras and Collins' approach~\cite{veras2019discriminability} by replacing MS-SSIM with deep-feature-based similarity metrics. One key difference between our approach and theirs is that while we exclusively rely on pre-trained ImageNet and Stylized ImageNet weights, Veras and Collins learned scaling weights for MS-SSIM using the scatterplots specific to their study. As such, their learned metric might not transfer well to different visualization types or domains. 

\begin{figure}[!htbp]
    \centering
    \includegraphics[width=\linewidth]{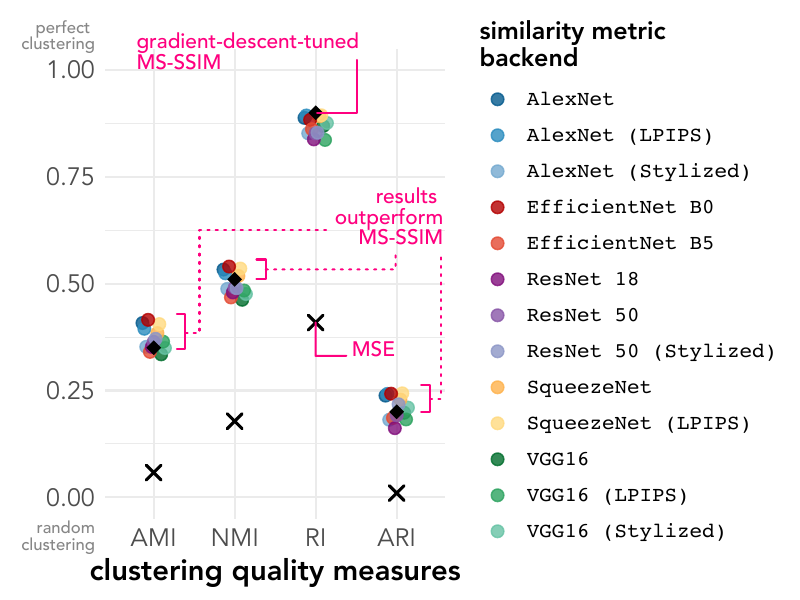}
    \caption{Black diamonds represent Veras and Collins' \cite{veras2019discriminability} results using gradient-descent-tuned MS-SSIM. Crosses represent results using Mean Squared Error (MSE). Points are offset for clarity.}
    \label{fig:pandey_compare}
    \Description{Some of the deep-feature-based similarity metrics, such as those that use the AlexNet architecture and the SqueezeNet architecture, have clustering quality measures better than the results obtained by Veras and Collins. All results out-perform results calculated using MSE.}
\end{figure}

For each ML network (\cref{tab:architecture_and_weights}), we calculate $\binom{247}{2} = 30, 381$ pairwise distances between all scatterplots and turn these pairwise distances into $247 \times 247$ distance matrices. For a baseline comparison, we also calculate pairwise distances using Mean Squared Error (MSE), a popular pixel-based metric. We then apply hierarchical clustering with Ward's agglomeration strategy to compute the clusters for each distance matrix,  using even-height tree cuts to yield 20 clusters --- matching the number used by Pandey et al.~\cite{pandey2016towards}. To evaluate performance, we calculate clustering quality measures between our labels and the labels obtained by Pandey et al.~\cite{pandey2016towards}. We also compare our results to the results obtained by Veras and Collins~\cite{veras2019discriminability} (represented by \includegraphics[height=\fontcharht\font`\B]{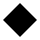} in \Cref{fig:pandey_compare}) and the baseline MSE results (represented by \includegraphics[height=\fontcharht\font`\B]{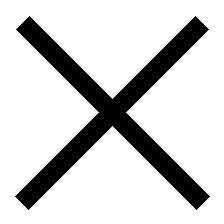} in \Cref{fig:pandey_compare}).

\subsection{Our Results}
\label{sec:pandey_results}

We find that our top-performing architectures ({\ColCircle{efftnetB0}}~\texttt{EfficientNet B0}, {\ColCircle{alex}} {\ColCircle{alexLPIPS}} {\ColCircle{alexSTY}} \texttt{AlexNet}, and {\ColCircle{squeeze}} {\ColCircle{squeezeLPIPS}} \texttt{SqueezeNet}) achieve comparable (if not superior) results to those achieved by gradient-descent-tuned MS-SSIM (\includegraphics[height=\fontcharht\font`\B]{06.png}), and far superior results to those achieved by MSE, the point-wise pixel difference metric. This outcome demonstrates robust performance across different CNN architectures and is particularly noteworthy as it relies purely on transfer learning with no training on the existing set of scatterplots. 

Comparing between architectures, we find that {\ColCircle{squeezeLPIPS}}~\texttt{SqueezeNet (LPIPS)} generally outperforms all the other networks.\footnote{{\ColCircle{squeezeLPIPS}}~\texttt{SqueezeNet (LPIPS)} appear most frequently in the top-3 performing similarity metric backends for all clustering quality measures.} Except for RI, the best-performing ML networks, {\ColCircle{efftnetB0}}~\texttt{EfficientNet B0}, {\ColCircle{alex}} \texttt{AlexNet}, and {\ColCircle{squeezeLPIPS}} \texttt{SqueezeNet (LPIPS)}, outperform the gradient-descent-tuned MS-SSIM results (\includegraphics[height=\fontcharht\font`\B]{06.png}) obtained in Veras and Collins~\cite{veras2019discriminability}.

Put differently, the clustering labels using only ImageNet pre-trained weights match human labels better than Veras and Collins' approach, which used parameters optimized for these scatterplots. Contrary to initial expectations, we find that incorporating stylized ImageNet weights does not consistently lead to performance improvements. For details, see the full set of clustered scatterplots, their labels (both from Pandey et al.~\cite{pandey2016towards} and from using {\ColCircle{squeezeLPIPS}} \texttt{SqueezeNet (LPIPS)}), and the table of all clustering performance measures (\cref{tab:freq}) in~\Cref{sec:clustering_plots}. 

\subsection{Ablation Study}
\label{sec:pandey_ablation}

To systematically analyze the factors contributing to the effectiveness of deep-feature-based similarity metrics, we conducted a series of ablation studies. The studies examine (1) the relative importance of network architecture versus learned weights, (2) the impact of different feature extraction layers on clustering quality, and (3) the role of supervised learning and dataset complexity in feature learning, comparing ImageNet-1K against simpler datasets.

\subsubsection{Robustness Check --- Random Weights}
\label{sec:pandey_robust}

\begin{figure}[!htbp]
    \centering
    \includegraphics[width=\linewidth]{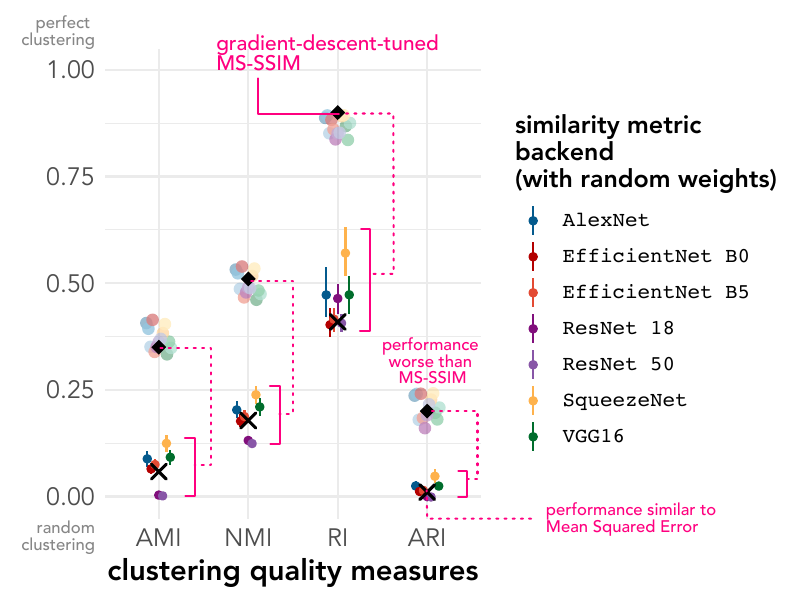}
    \caption{Black diamonds represent Veras and Collins'~\cite{veras2019discriminability} results using gradient-descent-tuned MS-SSIM. Crosses represent results using MSE. Dots show means of 10 trials. Lines indicate 95\% confidence intervals (CIs) from non-parametric bootstrap. Points and CIs are slightly offset for clarity.} 
    \label{fig:pandey_robust}
    \Description{The performance of deep-feature-based similarity metrics cluster around the results calculated using MSE.}
\end{figure}

We initialize the same architectures with \textit{random weights} to test whether our results stem from ImageNet pre-trained weights and assess if the resulting distance matrices and clustering labels still match empirical judgments. We repeat the process ten times and calculate the mean and 95\% non-parametric bootstrap confidence interval (CI) for each architecture (\cref{fig:pandey_robust}). 

We find that when using randomly initialized weights, all network architectures consistently \textit{under-perform} the results of the same networks that use ImageNet pre-trained weights (\cref{fig:pandey_compare}). In fact, the performance across network architectures is clustered around the performance of MSE, the pixel-based metric. These findings suggest that the clustering performance of deep-feature-based similarity metrics is largely attributable to weights trained on ImageNet.

\subsubsection{Impact of Each Feature Extraction Layer}

To identify which feature extraction layers contributed most to the clustering quality measures, we analyze the ``perceptual distance'' extracted from \textit{each individual layer} $l \in \mathcal{L}$, rather than summing across layers as in~\Cref{eq:distance_function}. We also examine the effect of each layer by computing distances using \textit{all layers except layer} $l \in \mathcal{L}$. We focus our analysis on the top-performing network architectures for each clustering quality measure.

\begin{figure}[!htbp]
    \centering
    \includegraphics[width=\linewidth]{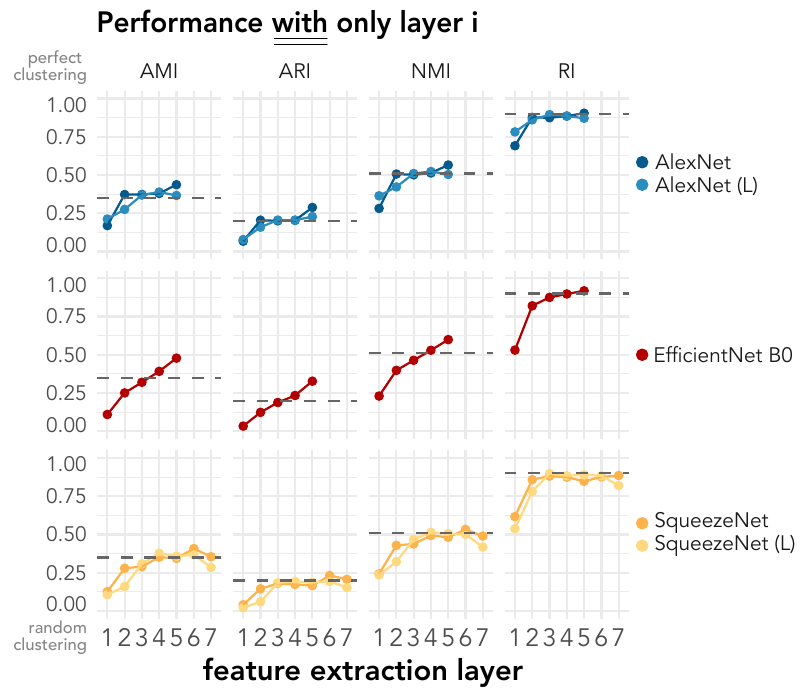}
    \caption[]{Clustering quality measures for {\ColCircle{alex}}~\texttt{AlexNet}, {\ColCircle{alexLPIPS}} \texttt{AlexNet (LPIPS)}, {\ColCircle{efftnetB0}} \texttt{EfficientNet B0}, {\ColCircle{squeeze}} \texttt{SqueezeNet}, and {\ColCircle{squeezeLPIPS}} \texttt{SqueezeNet (LPIPS)} calculated using perceptual distance at each layer $l \in \mathcal{L}$. Dashed lines (- - -) represent Veras and Collins' results using gradient-descent-tuned MS-SSIM.}
    \Description{First part of the first ablation study looking at clustering quality measures for the top-performing architectures. We see an increasing trend in performance when we go to the later layers.}
    \label{fig:pandey_ablation_sub1}
\end{figure}

Our analysis shows that across all examined neural network backbones, performance around layers 3 or 4 reaches performance comparable to those reported by Veras and Collins (\cref{fig:pandey_ablation_sub1}). The best per-layer performance is usually the last or second-to-last layer. 

\begin{figure}[!htbp]
    \centering
    \includegraphics[width=\linewidth]{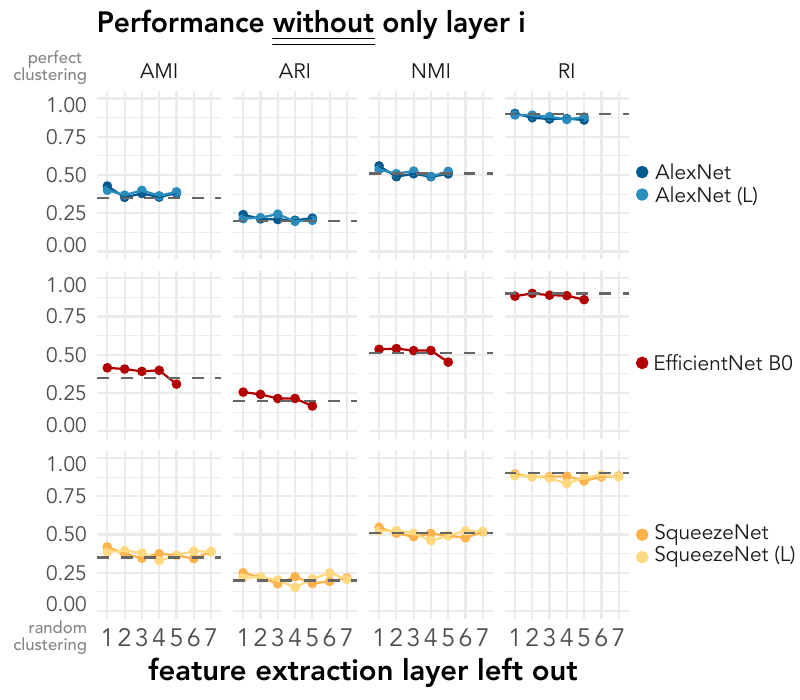}
    \caption[]{Clustering quality measures for {\ColCircle{alex}}~\texttt{AlexNet}, {\ColCircle{alexLPIPS}} \texttt{AlexNet (LPIPS)}, {\ColCircle{efftnetB0}} \texttt{EfficientNet B0}, {\ColCircle{squeeze}} \texttt{SqueezeNet}, and {\ColCircle{squeezeLPIPS}} \texttt{SqueezeNet (LPIPS)} calculated using all layers except layer $l \in \mathcal{L}$. Dashed lines (- - -) represent Veras and Collins' results using gradient-descent-tuned MS-SSIM.}
    \Description{Second part of the first ablation study looking at clustering quality measures for the top-performing architectures. The performance appears quite stable, except that if we take out the first layer, then performance in general slightly improves.}
    \label{fig:pandey_ablation_sub2}
\end{figure}

From \Cref{fig:pandey_ablation_sub2}, the performance appears relatively stable, with no specific layer significantly decreasing or increasing performance. The LPIPS version of \texttt{AlexNet} and \texttt{SqueezeNet} (i.e., ImageNet pre-trained weights + linear calibration, {\ColCircle{alexLPIPS}}~\texttt{AlexNet (L)} and {\ColCircle{squeezeLPIPS}}~\texttt{SqueezeNet (L)}) in general does not improve performance on top of the existing performance of {\ColCircle{alex}}~\texttt{AlexNet} and {\ColCircle{squeeze}}~\texttt{SqueezeNet}. These findings suggest that: (1) the linear calibration added by LPIPS provides limited value for scatterplot similarity; (2) while no single layer appears to dominate performance, excluding the first layer slightly improves performance; and (3) using only the last layer might be sufficient and could offer computational efficiency gains. These results align with prior observations~\cite{dingImageQualityAssessment2020, gatysNeuralAlgorithmArtistic2015} and show that later layers encode shape and structure information that better matches human similarity judgments. 

\subsubsection{Impact of Training Objective and Training Dataset} 
\label{sec:pandey-ablation-resnets}

We compared models pre-trained using different frameworks: supervised learning (ImageNet-1K and CIFAR-10~\cite{krizhevsky2009learning}) and self-supervised learning (SimCLR on STL-10~\cite{coates2011analysis}). Pre-trained weights were sourced from established repositories (see \cref{sec:neural_network_weights}).

\begin{figure}[!htbp]
    \centering
    \includegraphics[width=\linewidth]{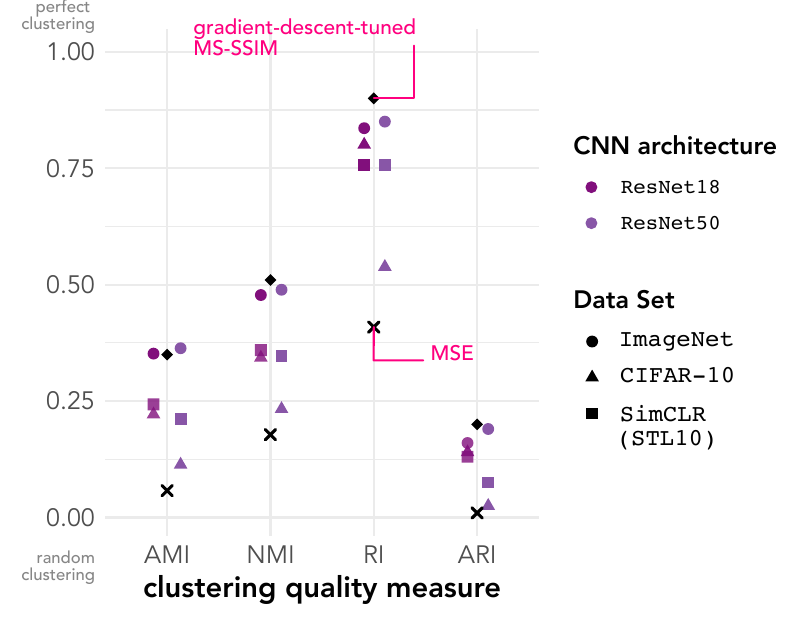}
    \caption[]{{\ColCircle{resnet18}} \texttt{Resnet18} and {\ColCircle{resnet50}} \texttt{ResNet50} performance. Black diamonds represent Veras and Collins'~\cite{veras2019discriminability} results using gradient-descent-tuned MS-SSIM. Crosses represent results using MSE. Points are dodged for clarity.}
    \Description{For both ResNet18 and ResNet50, the performance of architectures trained on CIFAR-10 and SimCLR (STIL10) is clearly not at the same level as those trained on ImageNet.}
    \label{fig:pandey_ablation_resnets}
\end{figure}

\Cref{fig:pandey_ablation_resnets} reveals a clear performance hierarchy: ImageNet-trained models performed best across all clustering metrics ({\ColCircle{resnet50}}~\texttt{ResNet50} slightly outperforming {\ColCircle{resnet18}}~\texttt{ResNet18}), followed by \btri ~CIFAR10-trained models, while SimCLR achieved similar results to \btri ~CIFAR-10 but underperformed compared to ImageNet. This gap could be due to differences in both the learning framework and dataset (STL-10 vs. ImageNet-1K).

We also observe an interesting pattern where {\ColCircle{resnet50}}~\texttt{ResNet50} trained on \btri~CIFAR10 performs similarly to the MSE baseline. This poor performance likely stems from the mismatch between {\ColCircle{resnet50}}~\texttt{ResNet50}'s deep architecture --- designed for complex, ImageNet-like datasets --- and CIFAR10's low-resolution ($32\times32$) images. With limited input complexity propagating through 50 layers, the network might default to learning shallow, pixel-level features rather than developing rich semantic representations. This demonstrates how dataset complexity and the learning framework impact representation quality and, in turn, downstream performance.

\subsection{Discussion of Veras and Collins}
\label{sec:pandey_discussion}

Our replication of Veras and Collins' study~\cite{veras2019discriminability} yields compelling results: without any training or fine-tuning on the stimuli, the best-performing deep-feature-based similarity metrics outperform gradient-descent-tuned MS-SSIM, a traditional computer vision metric, on three out of four clustering quality measures by an average of 14.04\% in aligning clustering labels to human labels of scatterplot. This is particularly noteworthy given that MS-SSIM's performance was specifically tuned on the scatterplots, while our deep learning models relied solely on pre-trained ImageNet weights. The ablation studies (\cref{sec:pandey_ablation}) suggest that (1) perceptual losses extracted at later stages are generally better (\cref{fig:pandey_ablation_sub1}) and (2) high-quality visual representations require both an appropriate learning framework and a sufficiently rich training dataset. The lower performance of CIFAR10-trained models (\cref{sec:pandey-ablation-resnets}), despite using supervised learning, emphasizes that dataset characteristics (resolution, class diversity, sample complexity) may be as important as the choice of learning framework. These results provide evidence for effective transfer learning from large-scale natural image datasets to data visualizations that utilize spatial encodings, suggesting that the features learned by neural networks on datasets such as ImageNet-1K may capture some fundamental aspects of visual perception that extend beyond the domain of natural images. 
\section[Replicating Demiralp et al. ]{Replicating Demiralp et al.~\cite{demiralp2014learning}}
\label{sec:demiralp}

\subsection{Overview of Original Paper}

\begin{figure}[!htbp]
    \centering
    \includegraphics[width=\linewidth]{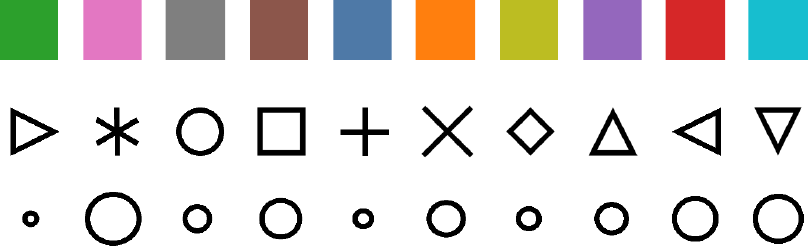}
    \caption{Palettes that maximize perceptual discriminability for color, shape, and size generated by Demiralp et al.~\cite{demiralp2014learning}, using the perceptual kernels from the triplet matching task.}
    \label{fig:palettes}
    \Description{Palette for color, from left to right: green, pink, gray, brown, blue, orange, yellow, purple, red, teal. Palette for shape, from left to right: rightward-pointing triangle, star, circle, square, plus sign, cross sign, diamond, upward-pointing triangle, leftward-pointing triangle, downward-pointing triangle. Palette for size, from left to right: size 1, size 10, size 4, size 7, size 2, size 6, size 3, size 5, size 8, and size 9.}
\end{figure}

Demiralp et al.~\cite{demiralp2014learning} introduced the concept of \textit{perceptual kernels}, which are distance matrices derived from aggregate perceptual judgments. Perceptual kernels encode perceptual differences in a reusable form that is directly applicable to visualization evaluation and automated design. For example, one can construct palettes that maximize perceptual discriminability, as shown in~\Cref{fig:palettes}. 

The authors conducted two sets of crowd-sourced experiments, one to estimate the perceptual kernels for \textit{color}, \textit{shape}, and \textit{size}, and one to estimate perceptual kernels for the pairwise combinations (i.e., \textit{color-shape}, \textit{color-size}, \textit{shape-size}). Each experiment used five different judgment types. They found that the triplet matching task, the task that asks participants to select between a pair of options that is most similar to the reference image, exhibits the least inter-subject variance, produces results that are less sensitive to subject count, and enables the most accurate prediction of bivariate kernels from univariate inputs \cite{demiralp2014learning}.  
\subsubsection{Data and Visual Stimuli} 
In their first experiment, Demiralp et al. \cite{demiralp2014learning} used the default color and shape palettes from Tableau, each of which contained ten distinct values. For size, they used a palette consisting of ten circles with linearly increasing area. In the second experiment, Demiralp et al. \cite{demiralp2014learning} selected four values from each palette and performed a cross product, resulting in $4 \times 4 = 16$ distinct values for each pairwise combination of color, shape, and size. Their source code, experiment interface, and data are at \url{https://github.com/uwdata/perceptual-kernels}.  

\subsubsection{Tasks}
Demiralp et al. elicited similarity judgments using five different tasks:

\begin{enumerate}[label=(\arabic*)]
    \item 
    \begin{minipage}[t]{\linewidth}
    \setlength{\intextsep}{0pt}
    \begin{wrapfigure}[4]{l}{0.32\linewidth}
        \centering
        \includegraphics[width=3cm]{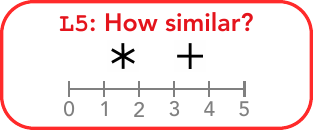}
    \end{wrapfigure}
    \textbf{Pairwise rating on a 5-point scale} (\task{l5}{L5}) presents participants with a pair of stimuli and asks them to rate the similarity between these two stimuli on a 5-point Likert scale.\footnotemark 
    \end{minipage}
    \footnotetext{These are actually 6-point Likert items (including 0), but we follow the naming scheme from the original work.}
    \item 
    \begin{minipage}[t]{\linewidth}
    \setlength{\intextsep}{0pt}
        \begin{wrapfigure}[4]{l}{0.32\linewidth}
          \centering
          \includegraphics[width=3cm]{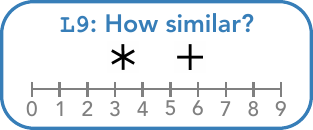}
    \end{wrapfigure}
    \textbf{Pairwise rating on a 9-point scale} (\task{l9}{L9}) presents participants with a pair of stimuli and asks them to rate the similarity between these two stimuli on a 9-point Likert scale.\footnotemark 
    \end{minipage}
    \footnotetext{Likewise, these are 10-point Likert items (including 0).}
    \item 
    \begin{minipage}[t]{\linewidth}
    \setlength{\intextsep}{0pt}
        \begin{wrapfigure}[4]{l}{0.32\linewidth}
        \centering
        \includegraphics[width=3cm]{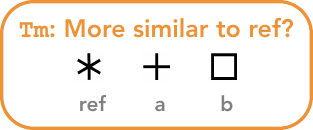}
        \end{wrapfigure}
        \textbf{Triplet ranking with matching} (\task{tm}{Tm}) presents participants with three stimuli, one of which is the reference, and asks participants to select among the two remaining options which one is most similar to the reference. 
    \end{minipage}
    \item 
    \begin{minipage}[t]{\linewidth}
    \setlength{\intextsep}{0pt}
        \begin{wrapfigure}[4]{l}{0.32\linewidth}
            \centering
            \includegraphics[width=3cm]{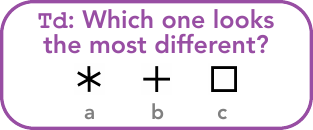}
        \end{wrapfigure}
    \textbf{Triplet ranking with discrimination} (\task{td}{Td}) asks participants to select the \textit{odd} stimuli out of a triplet of stimuli. This task is otherwise known as the \textit{odd-one-out} task.
    \end{minipage}
    \item 
    \begin{minipage}[t]{\linewidth}
    \setlength{\intextsep}{0pt}
        \begin{wrapfigure}[4]{l}{0.32\linewidth}
            \centering
            \includegraphics[width=3cm]{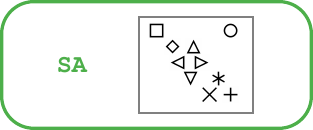}
        \end{wrapfigure}
    \textbf{Spatial arrangement} (\task{sa}{Sa}) presents participants with all stimuli and asks them to arrange them on a 2D plane.
    \end{minipage}
\end{enumerate}

A total of $6$ visual variables $\times \, 5$ judgment types $= 30$ jobs were run on Amazon Mechanical Turk, with each job completed by 20 Turkers for 600 distinct subjects \cite{demiralp2014learning}. 

\subsection{Implementation}

We replicate parts of Demiralp et al.'s~\cite{demiralp2014learning} experiments using deep-feature-based similarity metrics (\Cref{tab:architecture_and_weights}) to generate $10 \times 10$ and $16 \times 16$ distance matrices. For a baseline comparison, we also calculate pairwise distances using Mean Squared Error (MSE), a popular pixel-based metric. Because LAB color space is ``perceptually uniform''~\cite{cie1978recommendations}, meaning numerical distances between colors match human perception of those differences, we compute MSE in LAB rather than sRGB space for visual channels that involve color. Following Demiralp et al.'s approach, we normalize each matrix to span the range $[0, 1]$.

\subsubsection{Data and Visual Stimuli} 

Demiralp et al.'s experiments used stimuli of size $36 \times 36$, which is too small for certain deep-feature-based similarity metrics to operate on. Since these stimuli consist of elementary visual channels (single colors, shapes, and sizes) rather than complex data visualizations, we do not expect human perceptual judgments to be affected by increasing the stimulus size. Therefore, we generated 10 (16 for \textit{size-color}) PNG images in RGB color space of size $224 \times 224 \times 3$ for each visual channel. 

\subsubsection{Performance Evaluation}

Demiralp et al.~\cite{demiralp2014learning} compared the degree of compatibility between the five judgment tasks by calculating \textbf{Spearman's rank correlation coefficient} on pairwise distances for the same visual channel. Spearman's rank correlation coefficient, otherwise known as Spearman's $\rho$, is defined as the Pearson correlation coefficient between the variables' ranks and assesses how well the relationship between two variables can be described using a monotonic function. Spearman correlation ranges between $-1$ and $1$, with $1$ indicating perfect rank correlation and $-1$ indicating perfectly opposite ranks. This metric has also been used to evaluate the performance of computer vision similarity metrics, such as DISTS~\cite{dingImageQualityAssessment2020}, on Image Quality Assessment datasets. 

\subsection{Results}

For each of the four visual channels --- \textit{color}, \textit{size}, \textit{shape}, and \textit{size-color} --- we analyze the best-performing DL network.\footnote{... which is the one that achieves the highest average Spearman rank correlation coefficient when compared to Demiralp et al.'s perceptual kernels across their five judgment tasks.} For comparison, we show the perceptual kernel most correlated with the distance matrix of the best-performing DL network.\footnote{The complete set of distance matrices is available in the supplementary material.}

\begin{figure}[!htpb]
    \centering
    \includegraphics[width=0.75\linewidth]{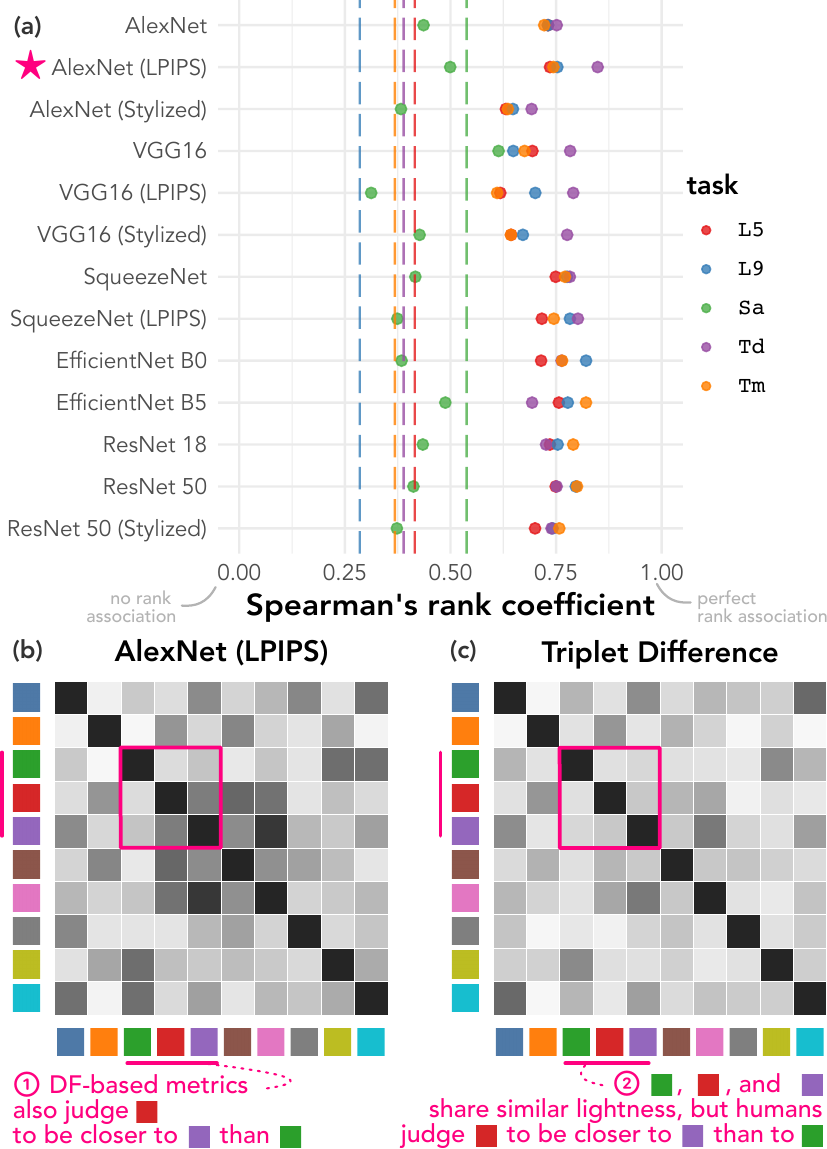}
    \caption{(a) Spearman's rank correlation between distance matrices and Demiralp et al.'s~\cite{demiralp2014learning} perceptual kernels (dashed: MSE-based distance matrices), (b) \texttt{AlexNet (LPIPS)} distance matrix, (c) Demiralp et al.'s \task{td}{Td} perceptual kernel.}
    \Description{Spearman rank correlations between deep-feature-based similarity metrics and perceptual kernels are lowest for spatial arrangement and highest for either triplet difference or triple matching. The patterns in matrix diagrams appear similar between AlexNet (LPIPS) and triplet difference.}
    \label{fig:demiralp_color_cka}
\end{figure}

\subsubsection{Color}
\label{sec:demiralp_color}

\texttt{AlexNet (LPIPS)} outperforms other DL networks. Distance matrices from deep-feature-based similarity metrics show the highest rank correlation with the perceptual kernel from \task{td}{Td} and the lowest with \task{sa}{Sa} (\cref{fig:demiralp_color_cka} top). 

\texttt{AlexNet (LPIPS)} demonstrates good overall performance in measuring relative color distances, but fails to replicate certain human perceptual patterns. For example, humans judge yellow to be more similar to green than teal, but \texttt{AlexNet (LPIPS)} ``sees'' yellow and teal as about equidistant from green.

The baseline MSE performances demonstrate the highest rank correlation with \task{sa}{Sa}, which makes sense given that MSE, a pixel-level metric, is directly measuring color differences in a perceptually uniform color space. This also suggests that \task{sa}{Sa} might be a better task to capture the non-linear aspects of human color perception. None of the pre-trained networks outperform the MSE baseline in \task{sa}{Sa}.

\subsubsection{Size}
\label{sec:demiralp_size}

\begin{figure}[!htpb]
    \centering
    \includegraphics[width=0.75\linewidth]{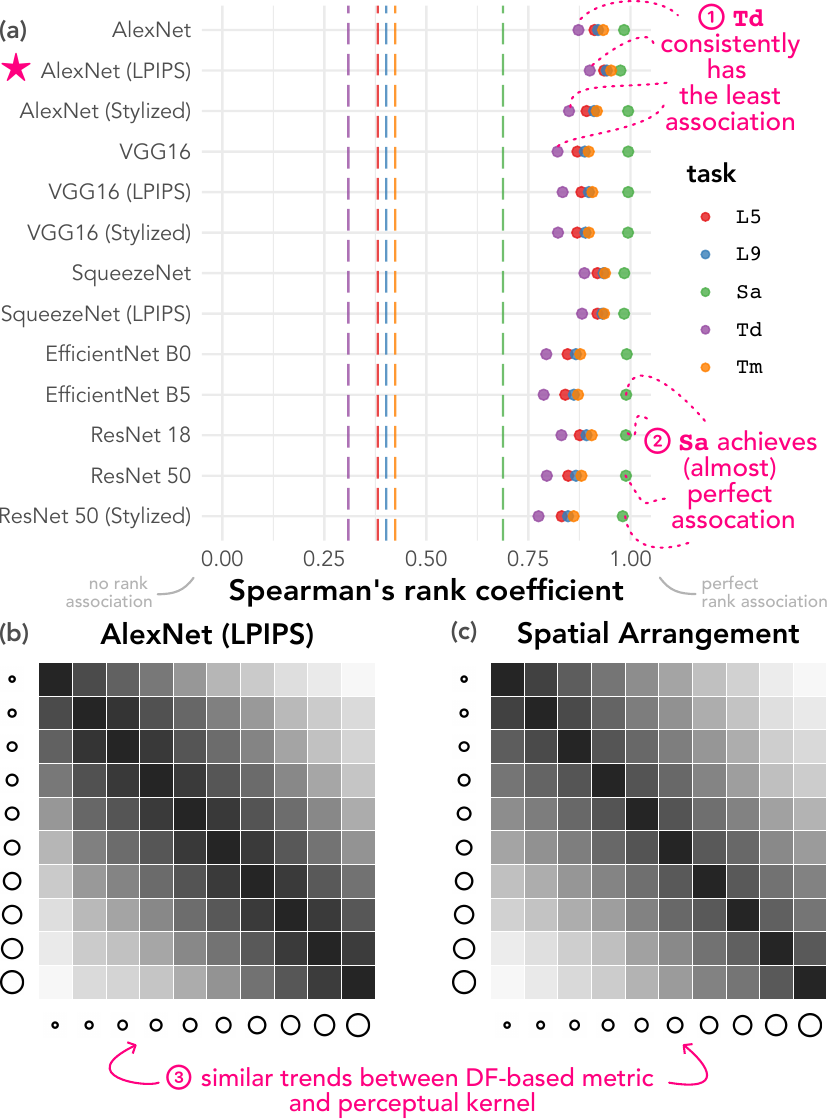}
    \caption{(a) Spearman's rank correlation between distance matrices and Demiralp et al.'s ~\cite{demiralp2014learning} perceptual kernels (dashed: MSE-based distance matrices), (b) \texttt{AlexNet (LPIPS)} distance matrix, (c) Demiralp et al.'s \task{sa}{Sa} perceptual kernel.}
    \Description{Spearman rank correlations between deep-feature-based similarity metrics and perceptual kernels are lowest for triplet difference and highest for spatial arrangement. The patterns in matrix diagrams appear very similar between AlexNet (LPIPS) and spatial arrangement.}
    \label{fig:demiralp_size_cka}
\end{figure}

\texttt{AlexNet (LPIPS)} outperforms other DL networks. Distance matrices from deep-feature-based similarity metrics and MSE show the highest rank correlation with the perceptual kernel from \task{sa}{Sa} and lowest with \task{td}{Td} (\cref{fig:demiralp_size_cka} (a)). 

Most DL networks achieve almost perfect rank correlation with \task{sa}{Sa} (\cref{fig:demiralp_size_cka} (a)) while MSE achieves around 0.7, suggesting these metrics capture more fundamental aspects of human size perception than pixel-level differences.

\subsubsection{Shape}
\label{sec:demiralp_shape}

\begin{figure}[!htpb]
    \centering
    \includegraphics[width=0.9\linewidth]{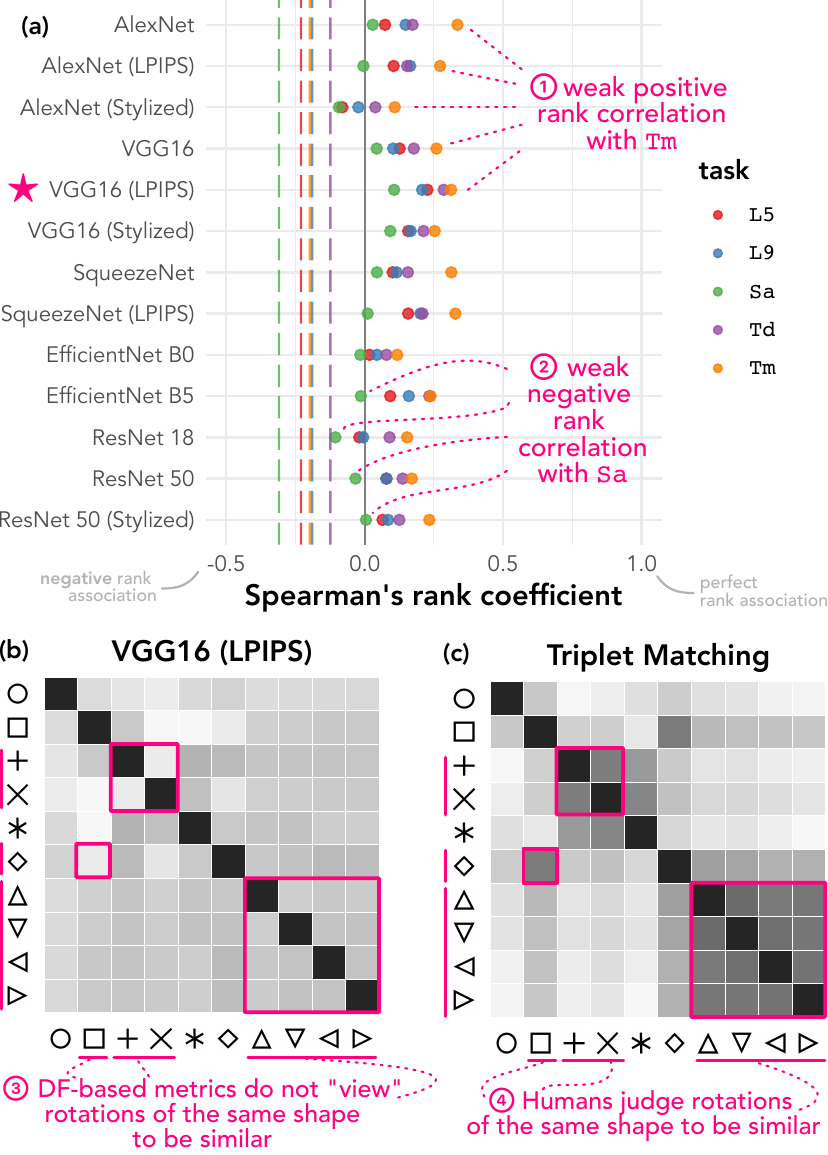}
    \caption{(a) Spearman's rank correlation between distance matrices and Demiralp et al.'s~\cite{demiralp2014learning} perceptual kernels (dashed: MSE-based distance matrices), (b) \texttt{VGG16 (LPIPS)} distance matrix, (c) Demiralp et al.'s \task{tm}{Tm} perceptual kernel.}
    \Description{Spearman rank correlations between deep-feature-based similarity metrics and perceptual kernels are lowest for spatial arrangement and highest for triplet matching. While correlations for earlier networks, such as AlexNet, VGG16, and SqueezeNet, appear weakly positive, the correlations for later networks, such as EfficientNet and ResNet, appear weakly negative. The patterns in matrix diagrams appear not at all similar between VGG16 (LPIPS) and Triplet Matching.}
    \label{fig:demiralp_shape_cka}
\end{figure}

\texttt{VGG16 (LPIPS)} outperforms other DL networks. Distance matrices from deep-feature-based similarity metrics show the highest rank correlation with the perceptual kernel from \task{tm}{Tm} and lowest with \task{sa}{Sa} (\cref{fig:demiralp_shape_cka} (a)). Stylized ImageNet weights weakly improve the correlation for \texttt{VGG16} and \texttt{ResNet50} but not \texttt{AlexNet}.

All MSE performances show a negative correlation, with \task{sa}{Sa} having the most negative correlation. Compared to MSE performances, the modest positive associations from CNNs show they are learning something, but there's still a large gap in capturing human shape perception. Specifically, all correlation coefficients are \textit{weak} (i.e., $\in$ [$-$0.25, 0.5]) compared to prior results for color (\cref{sec:demiralp_color} \cref{fig:demiralp_color_cka}) and size (\cref{sec:demiralp_size} \cref{fig:demiralp_size_cka}), where excluding \task{sa}{Sa}, rank correlations are around 0.75. Examining the distance matrix and perceptual kernel closely (\cref{fig:demiralp_shape_cka} (b), (c)) reveals that, unlike humans, deep-feature-based perceptual similarity metrics do not view similarity as \textbf{rotation-invariant}. In other words, while humans may judge \includegraphics[height=\fontcharht\font`\B]{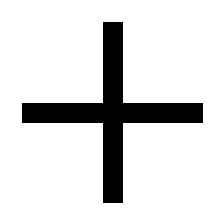} and \includegraphics[height=\fontcharht\font`\B]{cross_icon.png}, \includegraphics[height=\fontcharht\font`\B]{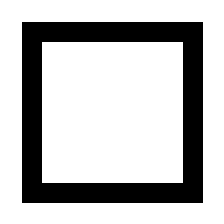} and \includegraphics[height=\fontcharht\font`\B]{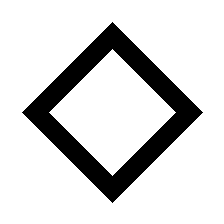} to be similar, as they are rotated variants of each other, \texttt{VGG16 (LPIPS)}, the best-performing DL network, ``sees'' them as different. We speculate why this is the case in \Cref{sec:demiralp_discussion}.

\subsubsection{Size-Color}
\label{sec:demiralp_size-color}

\begin{figure}[!htpb]
    \centering
    \includegraphics[width=0.9\linewidth]{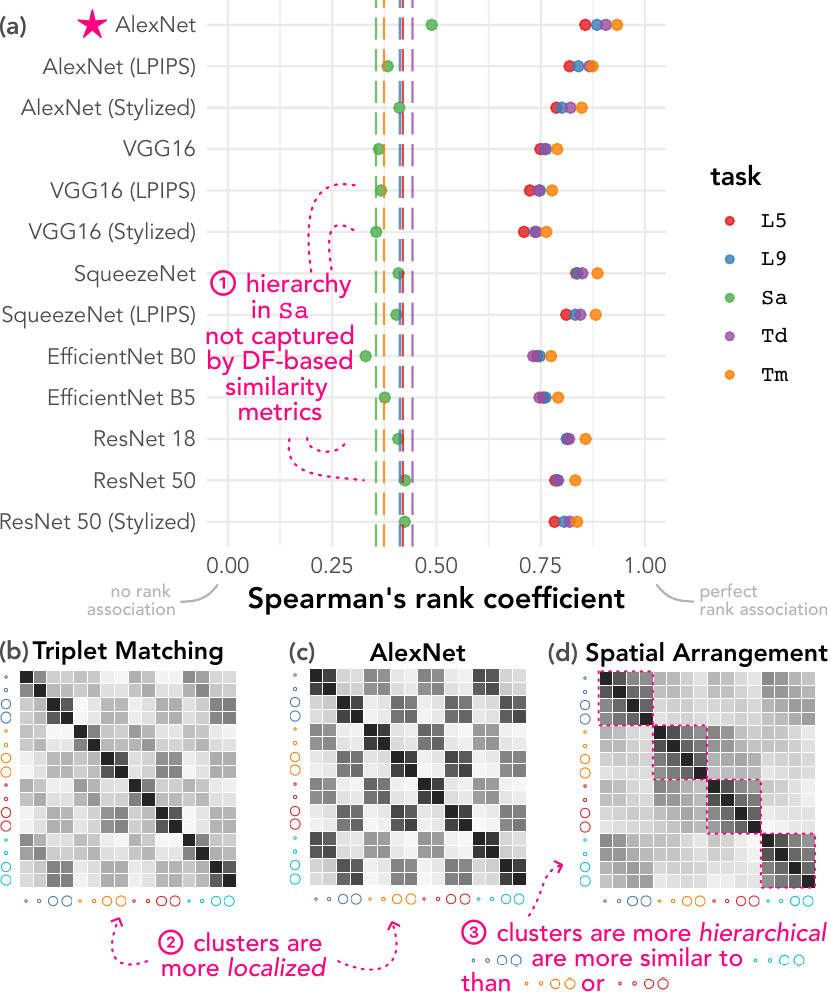}
    \caption{(a) Spearman's rank correlation between distance matrices and Demiralp et al.'s~\cite{demiralp2014learning} perceptual kernels (dashed: MSE-based distance matrices), (b) Demiralp et al.'s \task{tm}{Tm} perceptual kernel, (c) \texttt{AlexNet} distance matrix, (d) Demiralp et al.'s \task{sa}{Sa} perceptual kernel.}
    \Description{Spearman rank correlations between deep-feature-based similarity metrics and perceptual kernels are lowest for spatial arrangement and highest for triplet matching. The patterns in matrix diagrams appear similar between AlexNet and Triplet Matching, but not at all similar between AlexNet and Spatial Arrangement.}
    \label{fig:demiralp_size-color}
\end{figure}

\texttt{AlexNet} outperforms other DL networks. Distance matrices from deep-feature-based similarity metrics show the highest rank correlation with the perceptual kernel from \task{tm}{Tm} and the lowest with \task{sa}{Sa} (\cref{fig:demiralp_size-color} (a)). 

Interestingly, perceptual distances obtained via \task{sa}{Sa} (\cref{fig:demiralp_size-color} (d)) demonstrate a clear \textit{hierarchy} --- circles are most similar to those sharing the same color, followed by those with similar sizes, and then those with related colors (i.e., blue to teal, red to orange). 
In contrast, similarity judgments elicited through \task{tm}{Tm} (\cref{fig:demiralp_size-color} (b)) reveal more \textit{localized} similarity clusters, possibly due to the absence of \textit{context} and its implied constraints when using \task{tm}{Tm} to elicit similarity judgments. This suggests a potential need to reconsider Demiralp et al.'s advice to favor triplet matching (\task{tm}{Tm}) judgments \cite{demiralp2014learning}, especially when collecting visual variables that encode multiple visual channels simultaneously. 

This hierarchy of ``\textit{color} before \textit{size}'' is not captured by either MSE or deep-feature-based similarity metrics (\cref{fig:demiralp_size-color} (c)), and using ImageNet pre-trained weights in general does not improve correlation much against simple pixel-based MSE baseline.

\subsection{Robustness Check --- Random Weights}
\label{sec:demiralp_robust}

Similar to \cref{sec:pandey_robust}, we initialize the same architectures with random weights to determine to what extent ImageNet pre-trained weights can explain our results. We repeat the process of calculating distance matrices and Spearman's rank correlation coefficients with perceptual kernels ten times and present the 95\% non-parametric bootstrap CIs for Spearman's rank correlation coefficient. For each visual channel, we focus on the best-performing DL architecture and present the average distance matrix\footnote{All distance matrices can be found in the supplementary material.} that achieves the highest average Spearman correlation coefficient across all repetitions and all five judgment tasks. For direct comparison, we also present the perceptual kernel with which the average distance matrix has the highest Spearman correlation. 

\subsubsection{Color}

\begin{figure}[!htpb]
    \centering
    \includegraphics[width=0.9\linewidth]{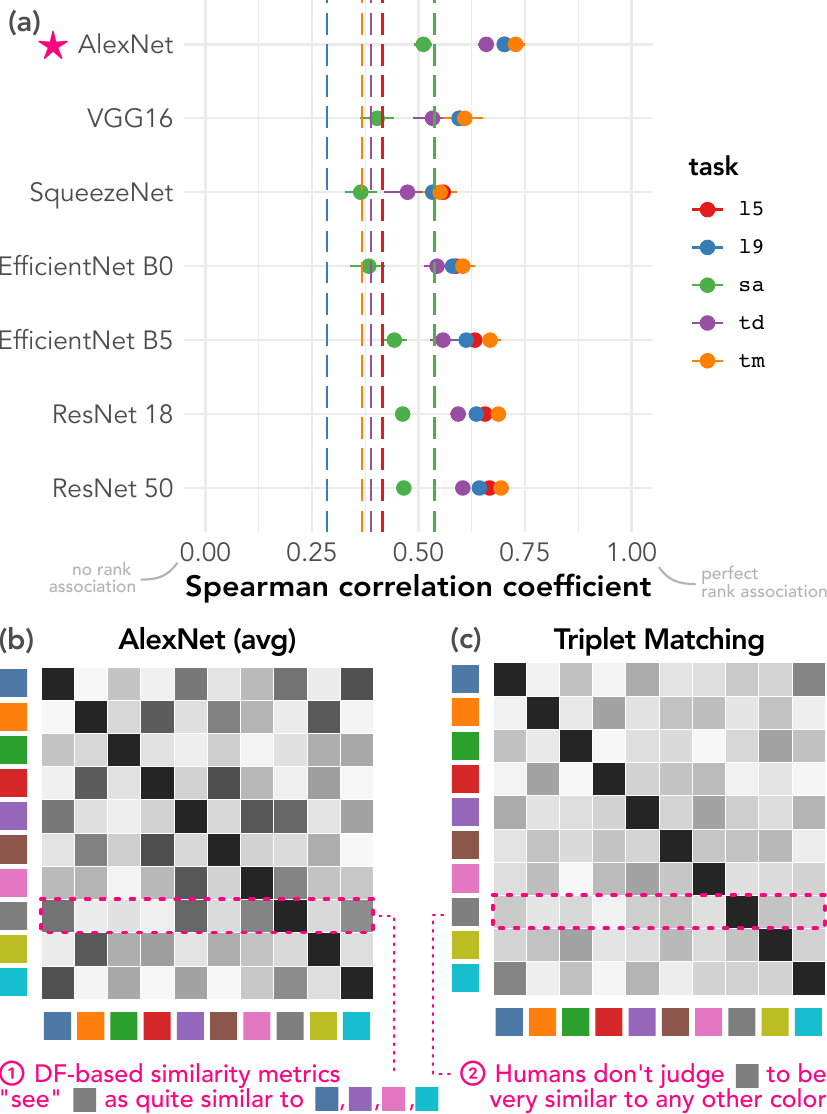}
    \caption{(a) Spearman's rank correlations between random-weight architectures' distance matrices and perceptual kernels (95\% bootstrap CI; dashed: MSE-based distance matrices), (b) Random \texttt{AlexNet}'s average distance matrix, (c) Demiralp et al.'s \task{tm}{Tm} perceptual kernel.}
    \label{fig:demiralp_color_robust}
    \Description{Spearman rank correlations between deep-feature-based similarity metrics and perceptual kernels are lowest for spatial arrangement and highest for triplet matching. The patterns in matrix diagrams appear very similar between AlexNet and triplet matching but not so much between AlexNet and spatial arrangement.}
\end{figure}

Correlation results with random weights (\cref{fig:demiralp_color_robust} (a)) are only slightly worse than those with ImageNet pre-trained weights (\cref{fig:demiralp_color_cka} (a)) in \cref{sec:demiralp_color}, suggesting that existing deep-feature-based similarity metrics may not effectively capture the perception of color similarity. 

\subsubsection{Size}

\begin{figure}[!htpb]
    \centering
    \includegraphics[width=0.9\linewidth]{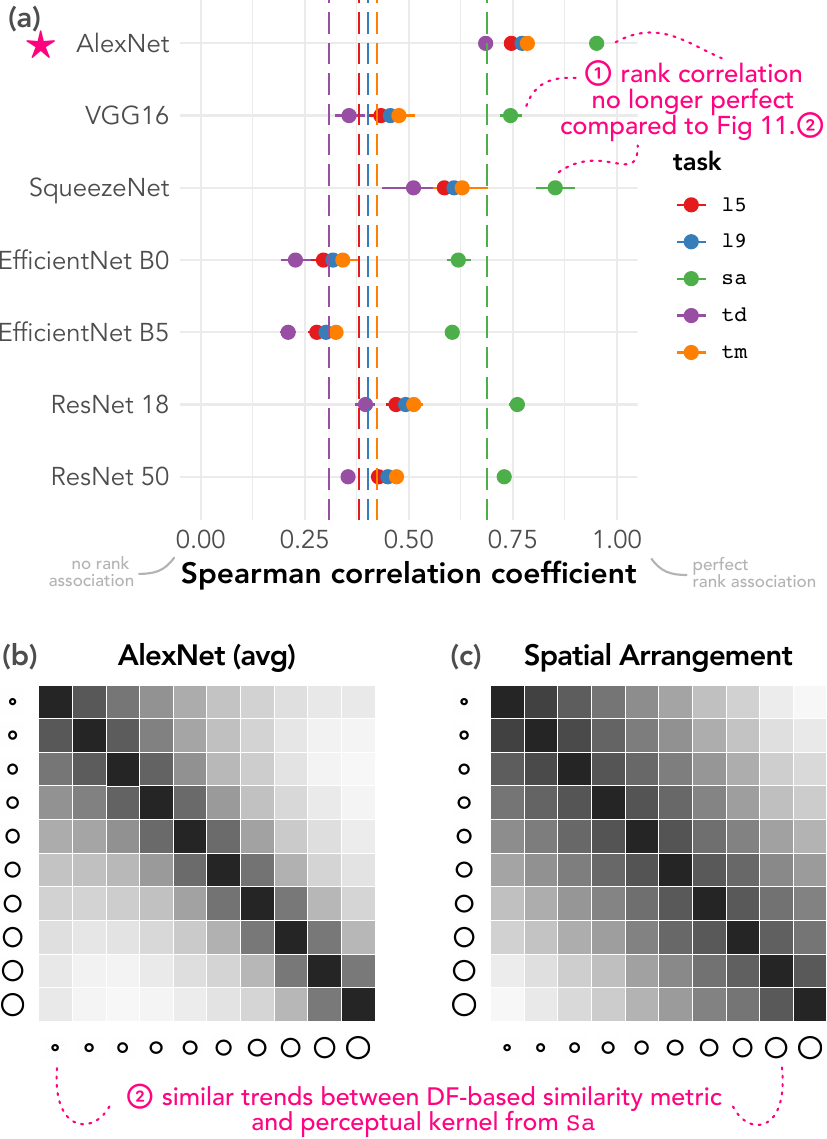}
    \caption{(a) Spearman's rank correlations between random-weight architectures' distance matrices and perceptual kernels (95\% bootstrap CI; dashed: MSE-based distance matrices), (b) Random \texttt{AlexNet}'s average distance matrix, (c) Demiralp et al.'s \task{sa}{Sa} perceptual kernel.}
    \label{fig:demiralp_size_robust}
    \Description{Spearman rank correlations between similarity metrics with randomly initialized weights and perceptual kernels are lowest for triplet difference and highest for spatial arrangement. The patterns in matrix diagrams appear similar between AlexNet (avg) and Spatial Arrangement.}
\end{figure}

The correlation results for size (\cref{fig:demiralp_size_robust} (a)), except for \texttt{AlexNet} and \texttt{SqueezeNet}, are now around 75\%, compared to perfect rank correlations for almost all networks (\cref{fig:demiralp_size_cka} (a)).

This suggests that deep-feature-based similarity metrics are learning something about size, but it is unclear to what extent this is attributable to ImageNet pre-trained weights or the functional form of deep-feature-based similarity metrics (\cref{eq:distance_function}) that relies on taking Euclidean differences between deep feature activations across spatial locations. For all tasks except \task{sa}{Sa}, using randomly initialized weights result in about similar rank correlation performance as using MSE (except for \texttt{AlexNet} and \texttt{SqueezeNet}).

\subsubsection{Shape}
\label{sec:demiralp_robust_shape}

\begin{figure}[!htpb]
    \centering
    \includegraphics[width=0.9\linewidth]{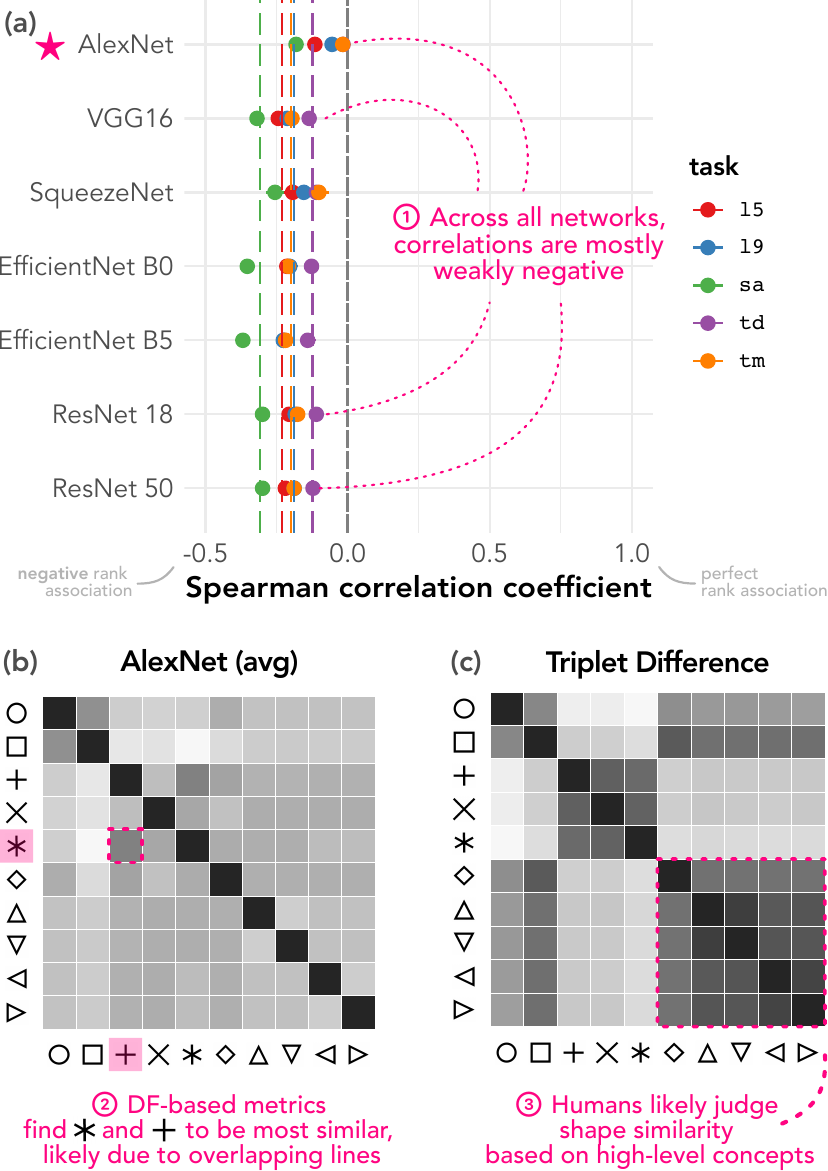}
    \caption{(a) Spearman’s rank correlations between random-weight architectures’ distance matrices and perceptual kernels (95\% bootstrap CI; dashed: MSE-based distance matrices), (b) Random \texttt{AlexNet}'s average distance matrix, (c) Demiralp et al.'s \task{td}{Td} perceptual kernel.}
    \label{fig:demiralp_shape_robust}
    \Description{Spearman rank correlations between similarity metrics with randomly initialized weights and perceptual kernels are lowest for spatial arrangement and highest for triplet difference, but they are all weakly negative. The patterns in matrix diagrams appear not at all similar between AlexNet (avg) and Tripet Difference.}
\end{figure}

Across all DL architectures, the performance is weakly negatively correlated to all perceptual kernels obtained via different tasks (\cref{fig:demiralp_shape_robust} (a)) and is worse than the previous correlation results (\cref{fig:demiralp_shape_cka} (a)) in \cref{sec:demiralp_shape}. 

The performance when using randomly initialized weights is very similar to that of using MSE. Given the existing functional form of deep-feature-based similarity metrics (\cref{eq:distance_function}) and its reliance on taking Euclidean differences between deep feature activations spatially, this result suggests that some transfer learning is happening in \Cref{fig:demiralp_size_cka}, but not to the degree that it captures what humans perceive as shape similarity. We also suspect that glyph shape similarity judgments are not entirely perceptual but may also depend on context/concept --- consider this triplet difference example: ($\times$, $\Box$, $\bigcirc$ ). One could choose $\bigcirc$ as the odd one out, since the other shapes have angles and circles do not have angles; or choose $\times$ as the odd one out, since both $\Box$ and $\bigcirc$ can be drawn on a page without lifting the pen off the page.

\subsubsection{Size-color}

\begin{figure}[!htpb]
    \centering
    \includegraphics[width=0.9\linewidth]{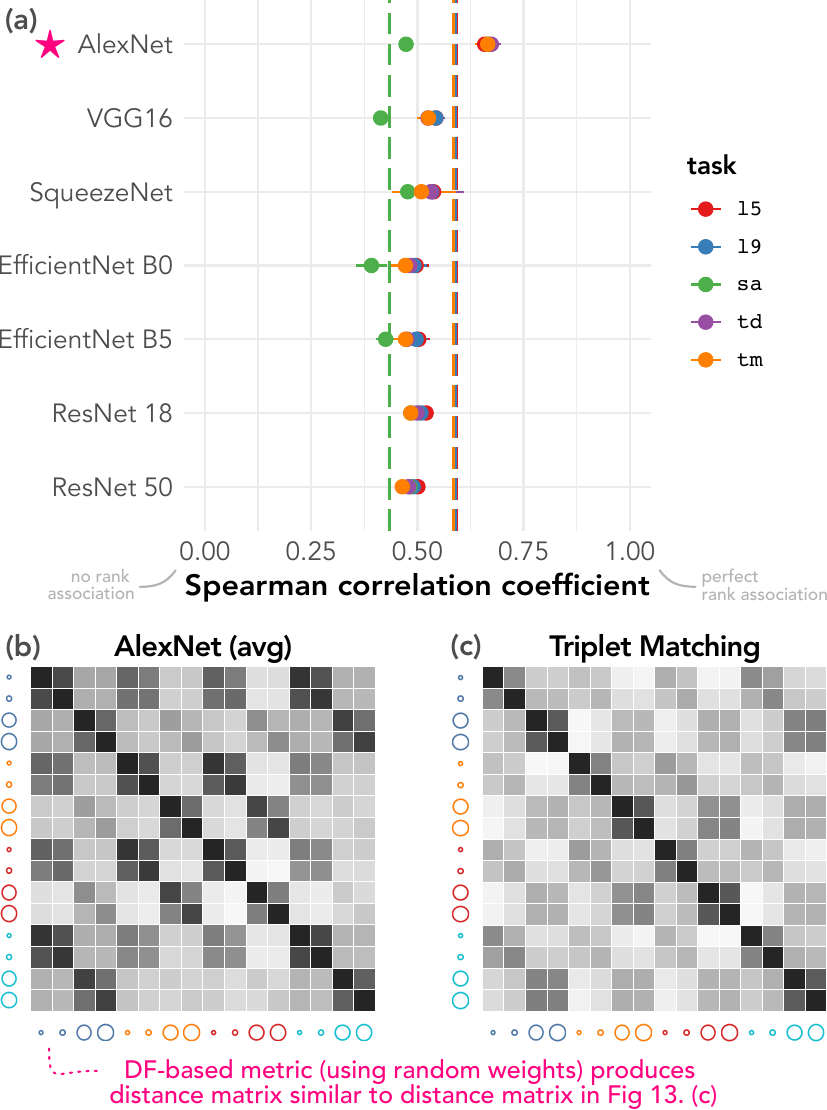}
    \caption{(a) Spearman’s rank correlations between random-weight architectures’ distance matrices and perceptual kernels (95\% bootstrap CI; dashed: MSE-based distance matrices), (b) Radom \texttt{AlexNet}'s average distance matrix, (c) Demiralp et al.'s \task{tm}{Tm} perceptual kernel.}
    \label{fig:demiralp_size-color_robust}
    \Description{Spearman rank correlations between similarity metrics with randomly initialized weights and perceptual kernels are lowest for spatial arrangement and highest for 5-point Likert. The patterns in matrix diagrams appear similar between AlexNet (avg) and Triplet Matching.}
\end{figure}

Again, we observe similar trends --- the rank correlation results (\cref{fig:demiralp_size-color_robust} (a)) are slightly worse compared to the rank correlation results (\cref{fig:demiralp_size-color} (a)) in \cref{sec:demiralp_size-color} when using ImageNet pre-trained weights. For all tasks, using randomly initialized weights result in rank correlations about the same as MSE.

Across the examined visual channels, the best-performing architecture, when using randomly initialized weight, is always \texttt{AlexNet}. We suspect that the simpler and shallower architecture and the larger filter size in the earlier layers ($11\times 11$, $5 \times 5$ instead of $3 \times 3$) of \texttt{AlexNet} might ``preserve'' more of the raw visual information, which explains its close to $0.5$ rank correlation with \task{sa}{Sa} for visual channels related to spatial information (\textit{size} and \textit{size-color}).

\subsection{Discussion of Demiralp et al.}
\label{sec:demiralp_discussion}

Although deep-feature-based similarity metrics achieve positive (\cref{fig:demiralp_color_cka,fig:demiralp_size-color}), if not perfect (\cref{fig:demiralp_size_cka}), rank correlation for several visual variables, the comparison against baseline MSE performances and robustness check (\cref{sec:demiralp_robust}) reveals that it is unclear whether such performance can be entirely attributed to pre-trained ImageNet or Stylized ImageNet weights. 

We suspect the decent correlation in the robustness check stems from both the functional form of deep-feature-based similarity metrics (\cref{eq:distance_function}) and Demiralp et al.'s ~\cite{demiralp2014learning} choice of visual stimuli. Deep-feature-based similarity metrics, as we implemented, calculate \textit{Euclidean differences between deep feature activations}, and for simple visual stimuli without complex patterns or textures, the feature map difference primarily reflects how well the stimuli \textit{spatially align/structurally correspond}. For example, with rotated shape stimuli (\includegraphics[height=\fontcharht\font`\B]{plus_icon.png} and \includegraphics[height=\fontcharht\font`\B]{cross_icon.png}), the feature map differences remain almost identical whether using pre-trained ImageNet or random weights, given the same DL architecture.  

Across all five similarity judgment tasks (\task{l5}{L5}, \task{l9}{L9}, \task{sa}{Sa}, \task{td}{Td}, and \task{tm}{Tm}), \task{sa}{Sa} shows the least correlation with deep-feature-based similarity metrics for all visual channels except \textit{size}, and has the least average rank correlation with all other perceptual kernels of the same visual channel (Demiralp et al.~\cite{demiralp2014learning}). We suspect this occurs because participants see the most visual stimuli simultaneously (all 10/16 of them), and \textit{therefore face more global, hierarchical constraints when judging their similarities}. In contrast, pairwise rating tasks (\task{l5}{L5}, \task{l9}{L9}) and triplet judgment tasks (\task{td}{Td}, \task{tm}{Tm}) only require focusing on the local relationships between 2--3 stimuli at a time. This provides one way to explain the \task{sa}{Sa} perceptual kernel for \textit{size-color} (\cref{fig:demiralp_size-color}), where participants judge similarity first by color, then by size within each same-color group. 

Our replication of Demiralp et al.~\cite{demiralp2014learning}'s experiment demonstrates that deep-feature-based similarity metrics currently \textit{do not approximate the perceptual distance between visual encodings} such as color and shape. Specifically, these metrics are \textit{not rotation-invariant} (\cref{sec:demiralp_shape}) and do not capture the \textit{precedence effect} of certain visual channels over others in similarity judgments (\cref{sec:demiralp_size-color}). Finally, while we explored using Stylized ImageNet weights, we cannot conclude definitively whether this approach improves performance, suggesting the need for further research.
\section{General Discussion}
\label{sec:discussion}

\subsection{Key Findings}
\begin{enumerate}
    \item \textbf{Effectiveness for complex visualizations}: Our replication of Veras and Collins' study \cite{veras2019discriminability} demonstrates that deep-feature-based similarity metrics can outperform traditional computer vision metrics (e.g., MS-SSIM) in aligning with human clustering judgments of scatterplot similarity. We highlight the fact that MS-SSIM parameters are optimized on the domain data (i.e., scatterplots) while deep-feature-based similarity metrics only rely on weights trained on \textit{natural images} and are therefore completely domain-free. This implies that transfer learning from natural image domains to visualizations is feasible and that ImageNet pre-trained weights capture fundamental visual features that generalize across diverse visual domains. This could be good in practice by significantly reducing the effort required to develop and train models from scratch for each new set of domain-specific visualizations. 
    \item \textbf{Limitations for abstract visual encodings}: In replicating Demiralp et al.'s work \cite{demiralp2014learning}, we find that deep-feature-based metrics struggle to capture human perceptual similarities for basic visual channels like color and glyph shape, but perform well when assessing size. We hypothesize that part of the reason for this poor performance is because judgments of color and glyph shape similarity are \textit{not purely perceptual}, that is, they also rely on high-level semantics and/or concepts. For example, different cultural associations with color hues ~\cite{jacobs1991cross} may lead people to judge the similarity of color between the same triplet differently. In terms of context/concept, see the example in \Cref{sec:demiralp_robust_shape}.
    We also observe that when participants use spatial arrangement to judge visual stimuli that encode multiple channels (e.g., \mbox{\cref{fig:demiralp_size-color}} in \mbox{\cref{sec:demiralp_size-color}}), they tend to judge color similarity \textit{before} size similarity.
    \item \textbf{Architecture and weight sensitivity:} Our results show that the performance of deep-feature-based metrics does not vary significantly across different neural network architectures (\cref{fig:pandey_compare}) but does vary across different pre-trained weights of varying complexity (e.g., CIFAR10 vs ImageNet-1K  in \Cref{fig:pandey_ablation_resnets}). The impact of stylizing ImageNet is less clear. 
\end{enumerate}

In summary, our findings suggest a potential for deep-feature-based similarity metrics in tasks involving visualizations that exhibit texture or pattern-like features (e.g., scatterplots) and highlight a need for caution when applying these metrics to primitive visualization glyphs or elements. 

\subsection{Limitations and Future work}

While deep-feature-based similarity metrics offer a promising solution for approximating similarity perception in information visualization, future work should explore the \textbf{generalizability} of deep-feature-based similarity metrics, focusing on:

\begin{enumerate}
    \item \textbf{Perceptual vs. conceptual similarity:} Our current approach focuses on low-level perceptual similarity and does not account for higher-level conceptual or semantic similarities \cite{pandey2016towards} that may be important in visualization interpretation.  
    \item \textbf{Visualizations are multi-modal and relational:} Our study primarily focuses on static, visual elements. An interesting direction for future work is to investigate how these metrics can be extended to handle different types of visualizations and their encoded visual relations~\cite{haehnEvaluatingGraphicalPerception2019, fleuret2011comparing, kim2018not}, diverse user groups, or multimodal visualizations that incorporate text, interactivity, or other non-visual elements. 
    \item \textbf{Benchmark development:} Deep-feature-based similarity metrics vary widely in their architectures~\cite{zhangUnreasonableEffectivenessDeep2018, kumarBetterImageNetClassifiers2022}, training/tuning datasets~\cite{dingImageQualityAssessment2020}, feature extraction layers~\cite{pihlgren2023systematic}, and evaluation approaches~\cite{sucholutsky2023getting}. Comprehensive benchmarks are needed to standardize the evaluation and comparison of these metrics across domains.
    \item \textbf{Explore more DL architectures, learning frameworks, and training datasets:} While this work explores five architectures (trained for image classification), future work can investigate Vision Transformers (ViT)~\cite{vaswani2017attention}, alternative learning frameworks (e.g., self-supervised learning), and low-bit quantization. However, adapting ViT presents a fundamental challenge because the Transformer architecture relies on fixed-size sequences of patches and positional encodings. For example, the ViT-B/16 model processes $224 \times 224$ images into 196 patches ($14 \times 14$), while our $64 \times 64$ resolution yields only 16 patches ($4 \times 4$). This significant mismatch in sequence lengths makes it impractical to transfer pre-trained ViT weights to our deep-feature-based similarity metric without developing novel positional encoding mappings --- an algorithmic challenge beyond this paper's scope.
    \item \textbf{Replication scope and inherited design choices:} By replicating prior studies, our work inherits their insights and methodological constraints in comparing human and machine behavior. While we show deep-feature-based metrics can fit into existing frameworks (as shown in \Cref{fig:teaser}), we are limited by these studies' choices in stimulus presentation, task design, and inference procedures. Future research should validate these metrics through new experimental designs that systematically vary how human similarity judgments are elicited, representations are inferred, and human-machine behaviors are compared.
\end{enumerate}

\subsection{Similarity, Constraints, and Inferred Representations}

Examining perceptual kernels from \textit{size-color} (\cref{sec:demiralp_size-color}) reveals that different similarity judgment tasks likely impose \textit{different constraints} on participants. These different constraints are then encoded in different outcome variables and translated into different inferred representations. Beyond traditional tasks such as triplet matching and triplet difference, there are also quadruplet matching~\cite{gogolou2018comparing}, octet matching~\cite{roads2017improving}, and the lineup task ~\cite{buja2009statistical}. Beyond the contextual constraints imposed by different similarity judgment tasks, human cognitive limitations also play a crucial role --- just as Hosseinpour et al. ~\cite{hosseinpour2024examining} demonstrated that increasing the number of frames in a small multiples visualization leads to a linear decline in accuracy, the complexity of similarity judgments may also tax our limited cognitive processing capacity.

Moreover, different visualization tasks may require different similarity models. Existing research has predominantly modeled similarity via a \textit{geometric model} \cite{torgerson1965multidimensional}. Such models assume minimality, symmetry, and triangle inequality. However, violations of all these assumptions have been empirically observed \cite{tversky1977features}. Further research is needed to understand how different judgment tasks affect behavioral outcomes, how task constraints influence representation inference, and how different similarity models perform across visualization contexts. 

\subsection{Potential Applications and Broader Implications}

While not yet a replacement for human testing, these metrics show promise for both pre- and post-experimental applications. For instance, they could help researchers narrow visualization design spaces before human studies by providing initial estimates of regions likely to be perceived as similar or dissimilar. They could also enable systematic comparison and evaluation of designs, such as complementing qualitative findings from human studies by quantifying how similar human-recreated charts are against the original charts they were shown~\cite{Proma2024}.

This work contributes to the broader discussion about the role of computational models in understanding and automating aspects of visualization design and evaluation \cite{demiralpVisualEmbeddingModel2014}. While our results show promise, they also highlight the complexity of human visual perception and the challenges in creating truly generalizable models. As the field progresses, interdisciplinary collaboration between visualization researchers, cognitive scientists, and machine learning experts will be crucial to developing more perceptually aligned computational models for visualization analysis and design.
\section{Conclusion}
\label{sec:conclusion}

In this paper, we explore the application of deep-feature-based similarity metrics to the domain of information visualization. Through two replication studies, we investigate how these metrics compare to traditional similarity measures and human judgments across scatterplots and different visual encodings. Our work demonstrates that deep features trained on diverse, large-scale natural image datasets (e.g., ImageNet-1K) transfer remarkably well to analyzing data visualizations like scatterplots, where spatial distributions are key. However, we observe limitations when applying these features to abstract visual primitives like glyph shapes or colors. This suggests that while deep features effectively capture the spatial and structural aspects of visualizations that encode data, they may be less suited for analyzing fundamental visual elements in isolation. These findings help delineate where deep perceptual metrics are most applicable in visualization analysis and design. By demonstrating both the potential and limitations of deep-feature-based similarity metrics, we contribute to the growing body of research at the intersection of machine learning and information visualization \cite{wu2021ai4vis} and provide valuable insights for researchers and practitioners seeking to develop more sophisticated deep-learning-based tools for visualization analysis and design.

\begin{acks}
We gratefully acknowledge Noah Shen for initiating the preliminary work that led to this project, though the research ultimately took a different direction. We thank Enrico Bertini for providing the 247 scatterplots originally used in Pandey et al.~\cite{pandey2016towards}, and Fumeng Yang for her invaluable feedback throughout our research process. This work would not have been possible without the publicly available data from Veras and Collins~\cite{veras2019discriminability} and Demiralp et al.~\cite{demiralp2014learning}, for which we are deeply appreciative. We also thank the members of the Mu Collective for their continued support and the reviewers for their constructive feedback. 
This work was partially supported by the National Science Foundation under award No. 1930642.
\end{acks}

\bibliographystyle{ACM-Reference-Format}
\bibliography{ref.bib}
\appendix 
\onecolumn

\section{Neural Network Architecture and Extraction Points}
\label{sec:extraction_points}

We followed the implementation of Zhang et al. \cite{zhangUnreasonableEffectivenessDeep2018} when extracting deep features from \texttt{AlexNet}, \texttt{SqueezeNet}, and \texttt{VGG16}. We followed the implementation of Kumar et al. \cite{kumarBetterImageNetClassifiers2022} when extracting deep features from \texttt{ResNets} and \texttt{EfficientNets}. See~\Cref{tab:extraction_points} for details. Other works, such as Pihlgren et al. \cite{pihlgren2023systematic}, have extracted four points for each network such that the chosen points represent ``early, semi-early, middle, and late'' layers in the convolutional layers. While Pihlgren et al.'s approach offers a more systematic way to sample features across network depth, we opted to align our extraction points with Zhang et al.~\cite{zhangUnreasonableEffectivenessDeep2018} and Kumar et al.~\cite{kumarBetterImageNetClassifiers2022} to enable direct comparisons, though it's worth noting that there is no broad consensus in the field regarding optimal feature extraction locations.

\begin{table*}[!htbp]
    \centering
    \caption{Network architecture, feature extraction points, ImageNet top-1 accuracy, and model size}
    \begin{tabular}{l l c c c} 
        \toprule 
        Network architecture & Feature extraction points & Top-1 accuracy (\%) & Model size (MB) & Num of Parameters \\ 
        \midrule 
        \texttt{AlexNet} & 1st, 2nd, 3rd, 4th, and 5th ReLU & 56.522 & 233.1 & 61, 100, 840\\ \hline 
        \texttt{VGG16} & 2nd, 4th, 7th, 10th, and 13th ReLU & 71.592 & 527.8 & 138, 357, 544 \\ \hline 
        \texttt{SqueezeNet} & \shortstack[l]{1st ReLU, \\ 2nd, 4th, 5th, 6th, 7th, and 8th Fire} & 58.178 & 4.7 & 1, 235, 496 \\ \hline 
         \shortstack[l]{\texttt{ResNet18}, \\ \texttt{ResNet50}} &  \shortstack[l]{1st Conv2d, 1st MaxPool2d, \\2nd, 3rd, and 4th Block Stack} & \shortstack[c]{69.758 \\ 76.13 } & \shortstack[c]{44.7 \\ 97.8 } & \shortstack[c]{11, 689, 512 \\ 25, 557, 032}\\\hline 
         \shortstack[l]{\texttt{EfficientNet B0}, \\ \texttt{EfficientNet B5}} & \shortstack[l]{1st Conv2d, \\ 2nd, 3rd, 4th, and 6th MBConv \\(mobile inverted bottleneck)} & \shortstack[c]{77.692 \\ 83.444} & \shortstack[c]{20.5 \\ 116.9 } & \shortstack[c]{5, 288, 548 \\ 30, 389, 784}\\ 
         \bottomrule
    \end{tabular}
    \label{tab:extraction_points}
\end{table*}

\section{Neural Network Weights}
\label{sec:neural_network_weights}

We obtained Stylized ImageNet weights from Geirhos et al.~\cite{geirhos2018imagenet}. Stylized ImageNet is generated by AdaIn style transfer~\cite{huang2017arbitrary} and the code is available at \url{https://github.com/rgeirhos/Stylized-ImageNet}. We utilize the Stylized ImageNet weights for \texttt{AlexNet}, \texttt{VGG16}, and \texttt{ResNet50} trained by Geirhos et al.~\cite{geirhos2018imagenet} at \url{https://github.com/rgeirhos/texture-vs-shape/}. For \texttt{ResNet-50},  Geirhos et al.~\cite{geirhos2018imagenet} used the standard \texttt{ResNet-50} architecture from \texttt{PyTorch}~\cite{paszke2019pytorch} (i.e., the \texttt{torchvision.models.resnet50} implementation). Geirhos et al.~\cite{geirhos2018imagenet} used batch size of 256 and trained on Stylized ImageNet for 60 epochs with Stochastic Gradient Descent (\texttt{torch.optim.SGD}) using a momentum term of 0.5, weight decay \texttt{1e-4} and a learning rate of 0.1 which was multiplied by a factor of 0.1 after 20 and 40 epochs of training. For \texttt{AlexNet} and \texttt{VGG16}, they used model architectures from \texttt{torchvision.models} and trained the networks under the identical circumstances as \texttt{ResNet-50}. Identical hyperparameter setting except for the learning rate --- the learning rate for \texttt{AlexNet} was set to 0.001 and for \texttt{VGG16} was 0.01 initially. Both learning rates were multiplied by 0.1 after 20 and 40 epochs of training (60 epochs in total).

We obtained CIFAR-10 weights \cite{krizhevsky2009learning} trained for \texttt{ResNet18} and \texttt{ResNet50} from \url{https://github.com/huyvnphan/PyTorch_CIFAR10}. We obtained the SimCLR weights \cite{chen2020simple} trained for \texttt{ResNet18} and \texttt{ResNet50} from \url{https://github.com/sthalles/SimCLR}. We use the same feature extraction points as specified in \cref{tab:extraction_points}.

\section[Pre-processing with resolution 224 x 224]{Pre-processing with Resolution $224 \times 224$}
\label{sec:process_224}

Similar to prior observations~\cite{kumarBetterImageNetClassifiers2022}, we also found that using standard ImageNet resolution ($224 \times 224$) yields slightly worse performance compared to a resolution of $64 \times 64$.

\begin{figure}[!htbp]
        \centering
        \includegraphics[width=0.5\linewidth]{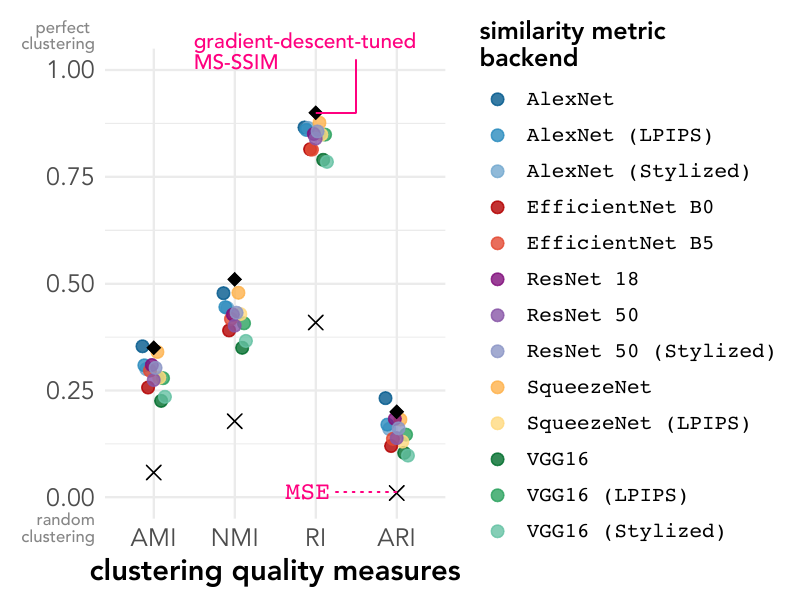}
        \caption{Black diamonds represent Veras and Collins' \cite{veras2019discriminability} results using gradient-descent-tuned MS-SSIM. Points are slightly offset for clarity. Results when replicating Pandey et al's experiment using resolution $224 \times 224 \times 3$.}
        \Description{The results from doing things at resolution 224 x 224.}
        \label{fig:pandey-results}
\end{figure}

\begin{table*}[!htbp]
  \caption{Clustering quality measures obtained from $64 \times 64$, highlighting the \textbf{\underline{top three}} performing backbones for each measure.}
  \label{tab:freq}
    \begin{tabular}{lrrrr}
    \toprule
    Backbone & RI & ARI & AMI & NMI\\
    \midrule
    Veras and Collins & 0.90 & 0.20 & 0.35 & 0.51 \\ 
    Mean Squared Error & 0.409 & 0.010 & 0.0579 & 0.178 \\ 
    \addlinespace
    \texttt{AlexNet} & 0.8878905 & 0.2378972 & \textbf{\underline{0.4078301}} & \textbf{\underline{0.5328779}}\\
    \texttt{AlexNet (Stylized ImageNet)} & 0.8517824 & 0.1814445 & 0.3521323 & 0.4876277\\
    \texttt{AlexNet (LPIPS)} & \textbf{\underline{0.8935190}} & \textbf{\underline{0.2419678}} & 0.3943937 & 0.5241815\\
    \addlinespace
    \texttt{VGG16} & 0.8688325 & 0.1978015 & 0.3345852 & 0.4622434\\
    \texttt{VGG16 (Stylized ImageNet)} & 0.8759751 & 0.2099710 & 0.3494193 & 0.4757726\\ 
    \texttt{VGG16 (LPIPS)} & 0.8363122 & 0.1822110 & 0.3643023 & 0.4839279\\
    \addlinespace
    \texttt{SqueezeNet} & \textbf{\underline{0.8917086}} & 0.2286150 & 0.3841792 & 0.5179834\\
    \texttt{SqueezeNet (LPIPS)} & \textbf{\underline{0.8939469}} & \textbf{\underline{0.2433410}} & \textbf{\underline{0.4053757}} & \textbf{\underline{0.5353521}}\\
    \addlinespace
    \texttt{EfficientNet B0} & 0.8833152 & \textbf{\underline{0.2424203}} & \textbf{\underline{0.4153032}} & \textbf{\underline{0.5400465}}\\
    \texttt{EfficientNet B5} & 0.8613278 & 0.1856533 & 0.3402646 & 0.4677815\\
    \addlinespace
    \texttt{ResNet 18} & 0.8375959 & 0.1617837 & 0.3518211 & 0.4793912\\
    \texttt{ResNet 50} & 0.8516507 & 0.1918074 & 0.3637146 & 0.4907449\\
    \texttt{ResNet 50 (Stylized ImageNet)} & 0.8531648 & 0.2176265 & 0.3707914 & 0.4893624\\
    \bottomrule
    \end{tabular}
\end{table*}

\section{Clustering of Scatterplots}
\label{sec:clustering_plots}

\begin{figure*}[!htbp]
    \centering
    \includegraphics[width=0.8\linewidth]{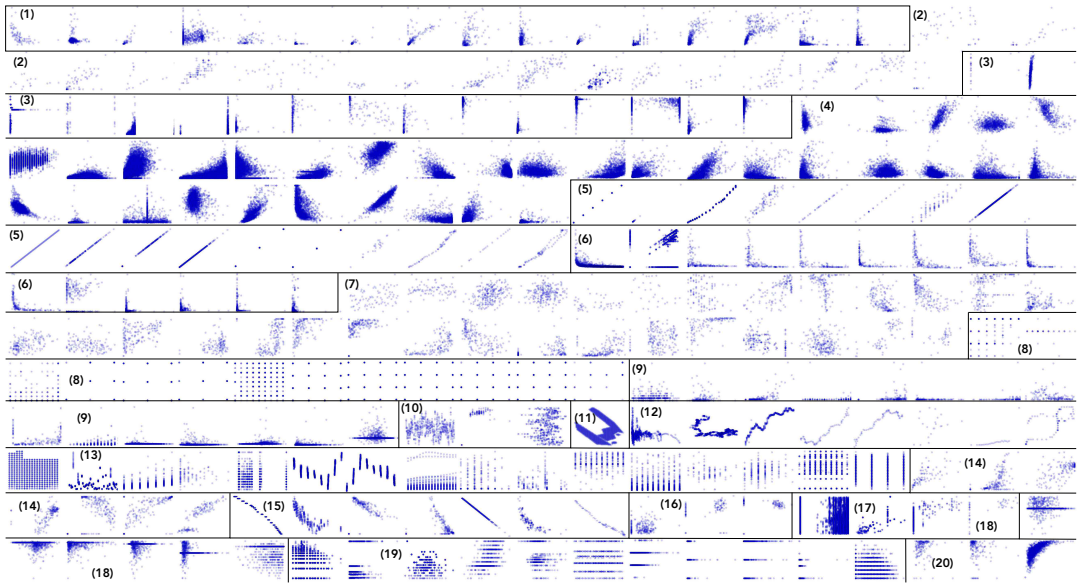}
    \caption{Consensus clustering of scatterplots from  Pandey et al.~\cite{pandey2016towards}.}
    \label{fig:pandey-cluster-results}
    \Description{Scatterplots clustered within the same group demonstrate visible similarity with each other.}
\end{figure*}

\begin{figure*}[!htbp]
    \centering
    \includegraphics[width=0.8\linewidth]{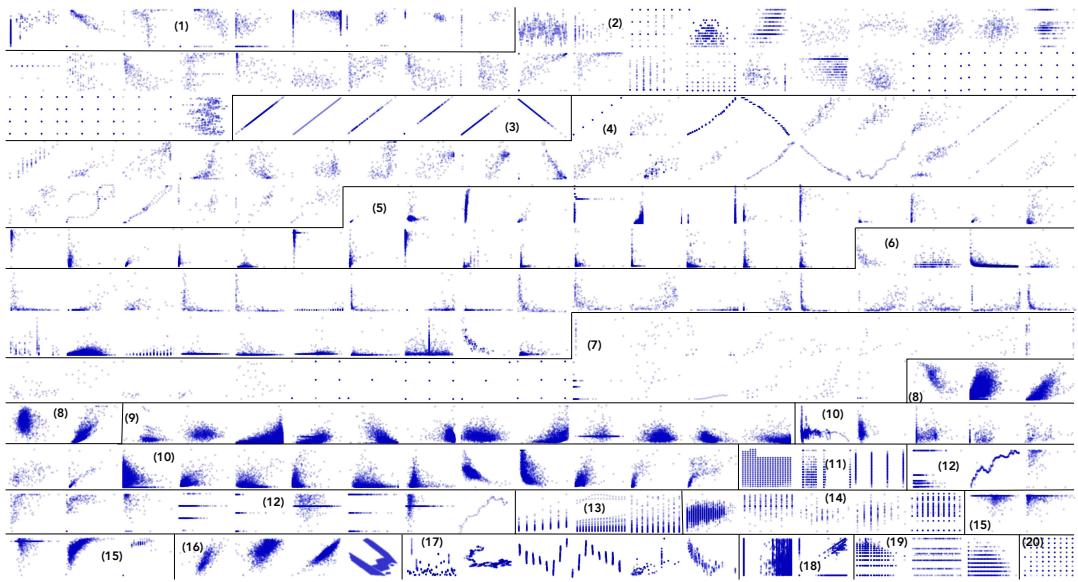}
    \caption{Clustering labels of scatterplots from Pandey et al. \cite{pandey2016towards} using \texttt{SqueezeNet (LPIPS)} as the backbone network for the deep-feature-based similarity metric.}
    \label{fig:custom-cluster-results}
    \Description{Scatterplots clustered within the same group demonstrate visible similarity with each other. However, the results don't quite agree with the clustering labels provided by Pandey et al. in Figure 19.}
\end{figure*}
\end{document}